\documentclass[12pt]{article}

\newcommand{\rau}{\partial}

\newcommand{\da}{\alpha}

\newcommand{\bango}[2]{\begin{equation}#1  \label{#2} \end{equation}}
\newcommand{\nn}{\nonumber}

\newcommand{\bra}[1]{\langle #1 |}
\newcommand{\ket}[1]{| #1 \rangle}
\newcommand{\bk}[2]{\langle #1 | #2 \rangle}

\newcommand{\eq}[1]{\begin{eqnarray} #1 \end{eqnarray}}
\newcommand{\ar}[4]{\left #1 \begin{array}{#2} #3 \end{array} \right#4}

\newcommand{\Maru}[1]{{\ooalign{\hfil#1\/\hfil\crcr\raise.167ex
\hbox{\mathhexbox20D}}}}

\begin{document}
%New environment
\newenvironment{Eqnarray}%
         {\arraycolsep 0.14em\begin{eqnarray}}{\end{eqnarray}}

\def\slash#1{{\rlap{$#1$} \thinspace/}}

\begin{flushright} 
%\today \\
January, 2001 \\
KEK-TH-737 \\
TIT/HEP-462 \\
%{\tt hep-th/00nnnnn}  
\end{flushright} 

\vspace{0.1cm}

\begin{Large}
       \vspace{0.1cm}
  \begin{center}
   {Noncommutative Gauge Theory on Fuzzy Sphere  \\
from Matrix Model}      \\
  \end{center}
\end{Large}
  \vspace{1cm}

\begin{center}
SATOSHI ISO $^{a)}$$^{c)}$
\footnote{e-mail address : satoshi.iso@kek.jp  }, 
YUSUKE KIMURA $^{a)}$$^{b)}$ 
\footnote{e-mail address : kimuray@post.kek.jp }, 
KANJI TANAKA $^{a)}$$^{c)}$ 
\footnote{e-mail address : kanji@post.kek.jp  }\\ 
and KAZUNORI WAKATSUKI $^{a)}$$^{d)}$
\footnote{e-mail address : waka@post.kek.jp  } \\
        
\vspace{0.5cm} 
$a)${\it  High Energy Accelerator Research Organization (KEK),}\\
{\it Tsukuba, Ibaraki 305-0801, Japan} \\
\vspace{0.2cm}   
$b)$ {\it  Department of Physics, Tokyo Institute of Technology, \\
Oh-okayama, Meguro-ku, Tokyo 152, Japan}\\
\vspace{0.2cm}
$c)${\it {\small Department of Particle and Nuclear Physics,
The Graduate University for Advanced Studies,}}\\
{\it {\small Tsukuba, Ibaraki 305-0801, Japan}} \\
\vspace{0.2cm}   
$d)${\it  Department of Physics, Tokyo Metropolitan University,}\\
{\it Hachiohji, Tokyo 192-0397, Japan} \\

\end{center}
        \vspace{0.6cm}

\begin{abstract}
\noindent
\end{abstract}

\noindent 
\hspace{0.4cm}
We derive a 
noncommutative $U(1)$ and $U(n)$ 
gauge theory on the fuzzy sphere
from a three dimensional 
matrix model 
by expanding the model around a classical solution of 
the fuzzy sphere. 
Chern-Simons term is added in the matrix model to make 
the fuzzy sphere as a classical solution of the model. 
Majorana mass term is also added to make it supersymmetric. 
We consider two large $N$ limits, 
one corresponding to a gauge theory on a commutative sphere and 
the other to that on a noncommutative plane. 
We also investigate stability of the fuzzy sphere by 
calculating one-loop effective action around classical solutions. 
In the final part of this paper, we consider another matrix model 
which gives a supersymmetric gauge theory on the fuzzy sphere. 
In this matrix model, only Chern-Simons term is added and 
supersymmetry transformation is modified.

\newpage 

\section{Introduction}
\hspace{0.4cm}
To formulate nonperturbative aspects of string theory or M theory, 
several kinds of matrix models 
have been proposed\cite{IKKT,BFSS}. 
These proposals are based on the  
developments of D-brane physics\cite{pol,Taylor}. 
IIB matrix model is one of these proposals\cite{IKKT}.
It is  a large $N$ reduced model\cite{reduced} 
of ten-dimensional supersymmetric 
Yang-Mills theory and 
the action has a matrix regularized form of 
the Green-Schwarz action of IIB superstring. 
It is postulated that it gives a constructive definition of 
type IIB superstring theory. 

In the matrix model, matter and even spacetime are dynamically emerged out 
of matrices\cite{AIKKT,AIKKTT}. 
Spacetime coordinates are represented  
by matrices and therefore noncommutative geometry appears naturally. 
The idea of the noncommutative geometry is to modify the microscopic 
structure of the spacetime. 
This modification is implemented by replacing fields 
on the spacetime by matrices. 
%The noncommutative structure remove the problem of 
%ultraviolet divergence in field theory 
%because of finite degrees of freedom of matrices\cite{Hoppe,madore}. 
It was shown\cite{AIIKKT,Li,BM,AMNS} that 
noncommutative Yang-Mills theories in a flat background  
are obtained by 
expanding the matrix model around a flat noncommutative background. 
The noncommutative background is a D-brane-like background 
which is a solution of the equation of motion and 
preserves a part of supersymmetry. 
Various properties of noncommutative Yang-Mills 
have been studied from the matrix model point of view
\cite{IIKK}.
%,12258,IKK,IIKK4038,AMNS2,11038}. 
In string theory, 
it is discussed that 
the world volume theory on D-branes 
with NS-NS two-form background 
is described by noncommutative Yang-Mills theory\cite{SW}.

%Recently, noncommutative Yang-Mills theory has been 
%studied in many situations. 
%It first appeared within the framework 
%of toroidal compactification of Matrix theory\cite{CDS}. 
%Much of the work in the noncommutative gauge theory 
%has been devoted to flat backgrounds which preserve a part supersymmetry. 

A different kind of noncommutative backgrounds, 
a noncommutative sphere, or a fuzzy sphere 
is also studied in many contexts. 
In \cite{Hoppe}, it is discussed in the framework of 
matrix regularization of a membrane. 
In the light-cone gauge, they gave
a map between functions on spherical membrane and hermitian matrices.
In BFSS matrix model\cite{BFSS}, membranes of spherical topology 
are considered in \cite{KabatTay,Rey,myers}. 
A noncommutative gauge theory on a fuzzy sphere 
in string theory context is discussed in \cite{ARS,ARS2}. 
The approach to construct a gauge theory on the fuzzy sphere 
were pursued in 
\cite{madore,GKP,WW,Klimcik,GM,GP, bala,9903112}. 

The fuzzy sphere\cite{madore} can be constructed 
by introducing a cut off parameter $N$ 
for angular momentum of the spherical harmonics. 
The number of independent functions is 
$\sum_{l=0}^{N}(2l+1)=(N+1)^{2}$. 
Therefore, 
we can replace the functions by $(N+1)\times (N+1)$ hermitian matrices 
on the fuzzy sphere. 
Thus, the algebra on the fuzzy sphere becomes noncommutative. 

In this paper, to construct a noncommutative gauge theory on the fuzzy sphere, 
we first consider a three dimensional supersymmetric reduced model.  
We add Chern-Simons term and Majorana spinor mass term 
to the action of original supersymmetric matrix model. 
An action with both of Yang-Mills and Chern-Simons terms
are also considered in \cite{ARS2}. 
Although an ordinary matrix model has only
a flat background as a classical solution, 
our matrix model can describe a curved background 
owing to these terms. 

We also study another type of action containing Chern-Simons term 
but not Majorana mass term. Although this action is not invariant 
under the original supersymmetry, it is invariant under 
a modified supersymmetry. It is also invariant 
under a constant shift of the fermion and hence it has 
${\cal N}=2$ supersymmetry. 

This paper is organized as follows. 
In section 2, we study a matrix model 
which gives a noncommutative gauge theory on a fuzzy sphere.  
We derive a noncommutative gauge theory 
by expanding the model around a classical solution 
which represents a fuzzy sphere. 
It is shown that in a commutative limit, 
it gives a standard theory on a commutative sphere. 
In the latter part of this section, 
another large $N$ limit corresponding to a sphere 
with a large radius  
but the noncommutativity fixed is considered. 
By taking the limit, 
we can obtain a noncommutative 
gauge theory in a flat noncommutative background, 
which is considered in \cite{AIIKKT}.  
In section 3, some properties of Dirac operator and 
chirality operator in this model are considered. 
These operators become the correct operators on a commutative sphere 
in a commutative limit. 
In section 4, stability of two clsssical solutions, 
diagonal matrices and the fuzzy sphere, is examined. 
One-loop effective action is calculated for the classical solutions. 
For diagonal matrices, all supersymmetry is preserved. 
Hence, one-loop effective action vanishes 
and eigenvalues can move freely. 
On the other hand, all supersymmetry is broken for the fuzzy sphere. 
Therefore one-loop effective action does not vanish. 
It is shown that the value of the action including the one-loop 
corrections of the fuzzy sphere is lower than that of the commuting 
matrices when $N$ is sufficiently large for fixed $g_{YM}$. 
Another matrix model is analyzed by the same manner as 
the case of a former model in section 5. 
While in the former model the fuzzy sphere does not preserve 
supersymmetry, in this model it is a BPS solution. 
Therefore we can obtain a supersymmetric noncommutative gauge theory 
on a fuzzy sphere  
by expanding this model around a fuzzy sphere. 
Section 6 is devoted to conclusions and discussions. 
In appendix A, a noncommutative product on a fuzzy sphere 
is constructed by following \cite{Hoppe}. 
In appendix B, by projecting 
a fuzzy sphere to {\it a fuzzy complex plane} 
stereographically 
using a coherent state approach developed by \cite{Bere,MMSZ,APS}, 
a noncommutative product which corresponds 
to the normal ordered product is considered. 
We also give 
mapping rules  from matrix models  to field theories on the 
projected plane.
This construction leads to field theories with
the Berezin type noncommutative product instead of 
the Moyal type.

% Let us comment about an operator ordering. 
% Matrices on the fuzzy sphere 
% are expanded by  noncommutative spherical harmonics. 
% The noncommutative spherical harmonics is constructed 
% by replacing commutative coordinates by noncommutative coordinates 
% in usual spherical harmonics 
% and same coefficient is used. 
% Ordering of the noncommutative coordinates 
% in the noncommutative spherical harmonics 
% corresponds to Weyl type ordering. 
% Therefore a product of the noncommutative spherical harmonics 
% becomes Moyal product in the decompactifying limit. 
% Stereographic projection from a fuzzy sphere 
% to a noncommutative complex plane 
% enable us to construct a normal ordered type basis. 
% After mapping from matrices to functions, 
% a product of the functions are written by Berezin product. 
% Construction of normal ordered product is summarized in Appendix B. 
% A map from matrices to functions are also given. 

%%%%%%%%%%%%%%%%%%%%%%%%%%%%%%%%%%%%%%%%%%%%%%%%%%%%%%

\section{Noncommutative Gauge Theory on Fuzzy Sphere}
\hspace{0.4cm}
We start with a three-dimensional 
%\footnote{Green-Schwarz string action is supersymmetric only when 
%spacetime dimenson is 3,4,6 and 10.} 
%large $N$ 
reduced model which is defined by the following action, 

\begin{equation}
 S= \frac{1}{g^{2}} Tr ( -\frac{1}{4} [ A_{i} ,A_{j}] 
         [A^{i} ,A^{j} ]  
  +\frac{2}{3}i\alpha \epsilon_{ijk}A^{i}A^{j}A^{k} 
  +\frac{1}{2}\bar{\psi}\sigma^{i}[A_{i},\psi] 
  +\alpha\bar{\psi}\psi ) .\label{action}
\end{equation}

\noindent 
Reduced models are obtained by reducing the spacetime volume 
to a single point\cite{reduced}.
We have added Chern-Simons term and Majorana mass term to 
the reduced model of supersymmetric Yang-Mills\cite{IKKT}. 
$A_{i}$ and $\psi$ are $(N+1) \times (N+1)$ hermitian matrices, 
and $\psi$ is a three-dimensional Majorana spinor field. 
$\sigma_{i}(i=1,2,3)$ denote Pauli matrices. 
$\alpha$ is a dimensionful parameter which depends on $N$. 

This model possesses $SO(3)$ symmetry and 
the following translation symmetry 
\begin{equation}
A_{i} \rightarrow A_{i}+c {\bf 1}. \label{translation}
\end{equation}
\noindent 
Gauge symmetry of this model is expressed 
by the unitary transformations,  
\begin{equation}
A_{i} \rightarrow UA_{i}U^{\dagger}, 
\hspace{0.3cm} \psi \rightarrow U\psi U^{\dagger}. \label{gaugetr}
\end{equation}
\noindent 
In addition to the above symmetries, 
this model has ${\cal N}=1$ supersymmetry:  
\begin{Eqnarray}
&&\delta A_{i} =i \bar{\epsilon}\sigma_{i}\psi \cr
&&\delta \psi  =\frac{i}{2}[A_{i} ,A_{j} ]
\sigma^{ij}\epsilon . \label{susy}
\end{Eqnarray}
\noindent 
Because of the fermionic mass term, 
translation symmetry of $\psi$ is not present. 

We next consider the equation of motion of (\ref{action}). 
When $\psi=0$, it is 
\begin{equation}
[A_{i},[A_{i},A_{j}]] =
-i\alpha\epsilon_{jkl}[A_{k},A_{l}].
\end{equation}
\noindent 
The simplest solution is realized by commuting diagonal matrices:
\begin{equation}
A_{i}= diag(x_{i}^{(N+1)},x_{i}^{(N)}, 
\cdots, x_{i}^{(1)}). \label{diagsol}
\end{equation}
% \noindent 
Another solution represents 
an algebra of the fuzzy sphere:  
\begin{equation}
[\hat{x}_{i},\hat{x}_{j}]=i\alpha\epsilon_{ijk}\hat{x}_{k}. 
\label{clsol}
\end{equation}
\noindent 
We impose the following condition for $\hat{x}_{i}$, 
\begin{equation}
\hat{x}_{1}\hat{x}_{1}+\hat{x}_{2}\hat{x}_{2}+\hat{x}_{3}\hat{x}_{3} =\rho^{2}.
\end{equation}
\noindent 
In the $\alpha \rightarrow 0$ limit, $\hat{x}_{i}$ becomes commutative 
coordinates $x_{i}$: 
\begin{Eqnarray}
x_{1}&=&\rho \sin \theta \cos \phi \cr
x_{2}&=&\rho \sin \theta \sin \phi \cr
x_{3}&=&\rho \cos \theta ,
\end{Eqnarray}
\noindent 
where $\rho$ denotes the radius of the sphere. 
Although the commuting matrices preserve the supersymmetry, 
the solution representing the fuzzy sphere breaks it. 
The noncommutative coordinates of (\ref{clsol}) can be constructed by 
the generators of the $(N+1)$-dimensional 
irreducible representation of $SU(2)$ 
\begin{equation}
\hat{x}_{i} =\alpha \hat{L}_{i}, \label{rep}
\end{equation}
where 
\begin{equation}
[\hat{L}_{i},\hat{L}_{j}]=i\epsilon_{ijk}\hat{L}_{k}. \label{su2}
\end{equation}
$\alpha$ and $\rho$ are related by the following relation
\begin{equation}
\rho^{2}=\alpha^{2}\frac{N(N+2)}{4}.
\end{equation}
\noindent 
where we have used the fact that the quadratic Casimir of $SU(2)$ 
in the $(N+1)$-dimensional irreducible representation is 
given by $N(N+2)/4$. 
The Plank constant, which represents the area occupied 
by the unit quantum on the fuzzy sphere, is given by 
\begin{equation}
\frac{4\pi \rho^{2}}{N+1}
= \frac{N(N+2)}{N+1}\pi\alpha^{2}. 
\end{equation}
\noindent 
Commutative limit is realized by 
\begin{equation}
\rho=\mbox{fixed},\hspace{0.4cm} N\rightarrow \infty 
\hspace{0.1cm}(\alpha \rightarrow 0). 
\label{comlim}
\end{equation}
\noindent 
From now on, we set $\rho$=fixed.

\vspace{0.5cm}

Now we show that, expanding the model around the classical backgrounds 
(\ref{clsol}) by the similar procedure as in \cite{AIIKKT,Li}
it leads to a noncommutative Yang-Mills on the fuzzy sphere.
We first consider $U(1)$ noncommutative gauge theory on the fuzzy sphere. 
We expand the bosonic matrices around the classical solution (\ref{clsol}):
\begin{equation}
A_{i}= \hat{x}_{i}+\alpha\rho\hat{a}_{i}=
\alpha \rho(\frac{\hat{L}_{i}}{{\rho}}+\hat{a}_{i}).
\label{bosonicmatexpansion}  
\end{equation}
\noindent 
A correspondence between matrices and functions on a sphere is given as follow.
Ordinary functions on the sphere can be expanded by the spherical harmonics, 
\begin{Eqnarray}
a(\Omega)&=&\sum_{l=0}^{\infty}\sum_{m=-l}^{l}
a_{lm}Y_{lm}(\Omega), \cr 
\psi(\Omega)&=&\sum_{l=0}^{\infty}\sum_{m=-l}^{l}
\psi_{lm}Y_{lm}(\Omega),  \label{function}
\end{Eqnarray}
\noindent 
where 
\begin{equation}
Y_{lm}= \rho^{-l}\sum_{a}f_{a_{1},a_{2},\cdots a_{l}}^{(lm)}
x^{a_{1}}\cdots x^{a_{l}} \label{harmonics}
\end{equation}
\noindent 
is a spherical harmonics and 
$f_{a_{1},a_{2},\cdots a_{l}}$ is a traceless and 
symmetric tensor. 
The traceless condition comes from $x_{i}x_{i}=\rho^{2}$. 
The normalization of the spherical harmonics is fixed by 
\begin{equation}
\int \frac{d\Omega}{4\pi} Y^{\ast}_{l^{\prime}m^{\prime}}Y_{lm}
=\frac{1}{4\pi} \int_{0}^{2\pi}d\varphi \int_{0}^{\pi} \sin \theta d\theta 
Y^{\ast}_{l^{\prime}m^{\prime}}Y_{lm}
=\delta_{l^{\prime}l}\delta_{m^{\prime}m} . \label{nor1}
\end{equation}
\noindent 
Matrices on the fuzzy sphere, on the other hand, 
can be expanded 
by {\sl the noncommutative spherical harmonics} $\hat{Y}_{lm}$ 
as 
\begin{Eqnarray}
\hat{a}=\sum_{l=0}^{N}\sum_{m=-l}^{l}
a_{lm}\hat{Y}_{lm}, \cr 
\hat{\psi}=\sum_{l=0}^{N}\sum_{m=-l}^{l}
\psi_{lm}\hat{Y}_{lm}.  \label{matrix}
\end{Eqnarray}
\noindent 
$\hat{Y}_{lm}$ is a $(N+1)\times(N+1)$ matrix and defined by 
\begin{equation}
\hat{Y}_{lm}=\rho^{-l}\sum_{a}f_{a_{1},a_{2},\cdots,a_{l}}^{lm}
\hat{x}^{a_{1}}\cdots\hat{x}^{a_{l}} , \label{nonspheri}
\end{equation}
\noindent 
where the same coefficients as (\ref{harmonics}) are used. 
Angular momentum $l$ is bounded at $l=N$ and these $\hat{Y}_{lm}$'s form 
a complete basis of $(N+1)\times(N+1)$ hermitian matrices.
From the symmetry of the indices, the ordering of $\hat{x}$ corresponds to 
the Weyl type ordering. 
A hermiticity condition requires that 
$a_{lm}^{\ast}=a_{l-m}$.
Normalization of the noncommutative spherical harmonics is given by  
\begin{equation}
\frac{1}{N+1}Tr(\hat{Y}^{\dagger}_{l^{\prime}m^{\prime}}
\hat{Y}_{lm})=\delta_{l^{\prime}l}\delta_{m^{\prime}m}. \label{normali}
\end{equation}
\noindent 
A map from matrices to functions is given by  
\begin{equation}
\hat{a}=\sum_{l=0}^{N}\sum_{m=-l}^{l}
a_{lm}\hat{Y}_{lm}
 \rightarrow 
a(\Omega)=\sum_{l=0}^{N}\sum_{m=-l}^{l}
a_{lm}Y_{lm}(\Omega),  
\end{equation}
\noindent 
and correspondingly 
a product of matrices is mapped to {\it the star product} on the 
fuzzy sphere:
\begin{equation}
\hat{a}\hat{b} \rightarrow a\star b. \label{productmap}
\end{equation}
\noindent 
Detailed structures of the star product are summarized 
in Appendix A.

An action of $Ad(\hat{L}_{3})$ is obtained by 
\begin{equation}
Ad(\hat{L}_{3})\hat{a}=\sum_{lm}a_{lm}[\hat{L}_{3},\hat{Y} _{lm} ] 
=\sum_{lm}a_{lm}m\hat{Y} _{lm}.
\end{equation}
\noindent 
This property and $SO(3)$ symmetry gives 
the following correspondence: 
\begin{Eqnarray}
Ad(\hat{L}_{i}) \rightarrow 
L_{i} \equiv \frac{1}{i}\epsilon_{ijk}x_{j}\partial_{k}. \label{angope}
\end{Eqnarray}
\noindent 
The laplacian on the fuzzy sphere is given by 
\begin{equation}
\frac{1}{\rho^{2}}Ad(\hat{L})^{2}\hat{a}
=\frac{1}{\rho^{2}}
\sum_{lm}a_{lm}[\hat{L}_{i}, [\hat{L}_{i},\hat{Y} _{lm} ]] 
=\sum_{lm}\frac{l(l+1)}{\rho^{2}}a_{lm}\hat{Y} _{lm}. 
\end{equation}
\noindent 
$Tr$ over matrices can be mapped to the integration over functions as 
\begin{equation}
\frac{1}{N+1}Tr \rightarrow \int \frac{d\Omega}{4\pi}. 
\end{equation}

\vspace{1cm}

Let us expand the action (\ref{action}) 
around the classical solution (\ref{clsol})
 and apply these mapping rules. 
The bosonic part of the action (\ref{action}) becomes 
\begin{Eqnarray}
S_{B}&=&-\frac{\alpha^{4}\rho^{4}}{4g^{2}}
Tr(\hat{F}_{ij}\hat{F}_{ij}) \cr 
&&-\frac{i}{2g^{2}}\alpha^{4}\rho^{3}\epsilon^{ijk}
Tr(\frac{1}{\rho}[\hat{L}_{i},\hat{a}_{j}]\hat{a}_{k}
+\frac{1}{3}\hat{a}_{i}[\hat{a}_{j},\hat{a}_{k}] 
-\frac{i}{2\rho}\epsilon_{ijm}\hat{a}^{m}\hat{a}_{k})
-\frac{\alpha^{2}}{6g^{2}}Tr({\hat{x}^{i}\hat{x}_{i}}) \cr
&=&
-\frac{\alpha^{4}\rho^{4}}{4g^{2}}(N+1)\int \frac{d\Omega}{4\pi}
(F_{ij}F_{ij})_{\star} \cr 
&&-\frac{i}{2g^{2}}\alpha^{4}\rho^{3}\epsilon^{ijk}
(N+1)\int \frac{d\Omega}{4\pi}
(\frac{1}{\rho}(L_{i}a_{j})a_{k}
+\frac{1}{3}a_{i}[a_{j},a_{k}] 
-\frac{i}{2\rho}\epsilon_{ijm}a^{m}a_{k})_{\star} \cr 
&&-\frac{\alpha^{4}}{24g^{2}}N(N+1)(N+2), \label{expandaction}
\end{Eqnarray}
\noindent where 
$\hat{F}_{ij}$ is defined as follows
\begin{Eqnarray}
\hat{F}_{ij}&=&\frac{1}{\alpha^{2}\rho^{2}} 
([A_{i},A_{j}] -i\alpha\epsilon_{ijk}A_{k} )\cr
&=&\frac{1}{\rho}[\hat{L}_{i},\hat{a}_{j}] 
-\frac{1}{\rho}[\hat{L}_{j},\hat{a}_{i}]
+[\hat{a}_{i},\hat{a}_{j}]-\frac{1}{\rho}i\epsilon_{ijk}\hat{a}_{k},  
\end{Eqnarray}
and the corresponding function $F_{ij}(\Omega)$ becomes
\begin{Eqnarray}
F_{ij}(\Omega)&=&\frac{1}{\rho}L_{i}a_{j}(\Omega)
 -\frac{1}{\rho}L_{j}a_{i}(\Omega)
+[a_{i}(\Omega),a_{j}(\Omega)]_{\star}
-\frac{1}{\rho}i\epsilon_{ijk}a_{k}(\Omega).  
\label{Fieldstrength}
\end{Eqnarray}
This quantity is gauge covariant and 
becomes zero when the fluctuation is set to zero. 
The Yang-Mills coupling is found to be 
$g_{YM}^{2}=4\pi g^{2}/(N+1)\alpha^{4}\rho^{2}$. 
%To obtain a correct commutative limit, 
%we need to fix $g_{YM}$. 
$(\hspace{0.2cm})_{\star}$ means 
that the products should be taken as the star product. 
Hence commutators do not vanish even in the case of 
$U(1)$ gauge group. 

The fermionic part of the action becomes 
\begin{Eqnarray}
S_{F}&=&\frac{\alpha}{2g^{2}}Tr(\bar{\hat{\psi}}\sigma^{i}
  [\hat{L}_{i}+\rho\hat{a}_{i},\hat{\psi}] 
+2\bar{\hat{\psi}}\hat{\psi}) \cr 
&=& 
\frac{\alpha\rho}{2g^{2}}(N+1)
\int \frac{d\Omega}{4\pi} 
(\frac{1}{\rho}\bar{\psi}\sigma^{i}L_{i}\psi 
+\bar{\psi}\sigma^{i}[a_{i},\psi] +\frac{2}{\rho}\bar{\psi}\psi)_{\star}. 
%\cr 
%&=& 
%\frac{\alpha\rho}{2g^{2}}(N+1) \int \frac{d\Omega}{4\pi}
%(\bar{\psi}D\psi+\bar{\psi}\sigma^{i}[a_{i},\psi] 
%+\frac{1}{\rho}\bar{\psi}\psi)_{\star} ,
\end{Eqnarray}
%\noindent 
%where $D=\frac{1}{\rho}(\sigma^{i}L_{i}+1)$ is a Dirac operator 
%on a sphere
%(Details are discussed in the next section). 

\vspace{0.3cm}

We next focus on the gauge symmetry of this action. 
The action (\ref{action}) is invariant under 
the unitary transformation (\ref{gaugetr}).  
For an infinitesimal transformation 
$U=\exp(i\hat{\lambda}) \sim 1+i\hat{\lambda}$ in (\ref{gaugetr}) 
where $\hat{\lambda}=\sum_{lm}\lambda_{lm}\hat{Y}_{lm}$, 
the fluctuation around the fixed background 
transforms as 
\begin{Eqnarray}
\hat{a}_{i} &\rightarrow& \hat{a}_{i} 
-\frac{i}{\rho}[\hat{L}_{i},\hat{\lambda}]
  +i[\hat{\lambda},\hat{a}_{i}] .
\label{matrixgauge}  
\end{Eqnarray}
After mapping to functions, we have local gauge symmetry
\begin{Eqnarray}
%\hat{a}_{i} &\rightarrow& \hat{a}_{i} -i[\hat{L}_{i},\hat{\lambda}]
%  +i[\hat{\lambda},\hat{a}_{i}] \cr 
a_{i} &\rightarrow& a_{i} -\frac{i}{\rho}L_{i}\lambda
  +i[\lambda,a_{i}]_{\star} .  \label{gatr}
\end{Eqnarray}

Let us discuss a scalar field which 
is defined by
\begin{Eqnarray}
\hat{\phi}&\equiv&\frac{1}{2\alpha \rho}
(A_{i}A_{i}- \hat{x}_{i}\hat{x}_{i}) \cr 
&=&\frac{1}{2}
(\hat{x}_{i}\hat{a}_{i}+\hat{a}_{i}\hat{x}_{i} 
+\alpha\rho \hat{a}_{i}\hat{a}_{i}).  
\end{Eqnarray}
\noindent 
It transforms covariantly as an adjoint representation
\begin{equation}
\hat{\phi} \rightarrow \hat{\phi} +i[\hat{\lambda}, \hat{\phi}].
\end{equation}

\noindent 
Since the scalar field should become 
the radial component of $\hat{a}_{i}$ in the commutative limit, 
a naive choice is
$\hat{\phi}_{0}=
(\hat{x}_{i}\hat{a}_{i}+\hat{a}_{i}\hat{x}_{i})/2$. 
For small fluctuations this field is 
the correct component of the field $\hat{a}_{i}$ but 
 large fluctuations of $\hat{a}_{i}$ deform the shape 
of the sphere and $\hat{\phi}_{0}$ can be no longer interpreted 
as the radial component of $\hat{a}_{i}$. 
This is a manifestation of the fact that matrix models or 
noncommutative gauge theories naturally unify spacetime and 
matter on the same footing.
An addition of the non-linear term 
$\hat{a}_{i}\hat{a}_{i}$ makes $\hat{\phi}$ transform correctly 
as the scalar field in the adjoint representation. 

\vspace{0.4cm}

We have so far discussed the $U(1)$ noncommutative gauge theory 
on the fuzzy sphere. 
A generalization to $U(m)$ gauge group is realized 
by the following replacement: 
\begin{equation}
\hat{x}_{i} \rightarrow \hat{x}_{i}\otimes{\bf 1}_{m}. 
\end{equation}
\noindent 
$\hat{a}$ is also replaced as follows:
\begin{equation}
\hat{a} \rightarrow 
\sum_{a=1}^{m^{2}}\hat{a}^{a}\otimes T^{a}, 
\end{equation}
\noindent 
where $T^{a} (a=1,\cdots, m^{2})$ denote the generators of $U(m)$.
Then we obtain $U(m)$ noncommutative gauge theory 
by the same procedure as the $U(1)$ case: 
\begin{Eqnarray}
S_{B}&=&
-\frac{\alpha^{4}\rho^{4}}{4g^{2}}(N+1)tr\int \frac{d\Omega}{4\pi}
(F_{ij}F_{ij})_{\star} \cr 
&&-\frac{i}{2g^{2}}\alpha^{4}\rho^{3}\epsilon^{ijk}
(N+1)tr\int \frac{d\Omega}{4\pi}
(\frac{1}{\rho}(L_{i}a_{j})a_{k}
+\frac{1}{3}a_{i}[a_{j},a_{k}] 
-\frac{i}{2\rho}\epsilon_{ijm}a^{m}a_{k})_{\star} \cr 
&&-\frac{\alpha^{4}}{24g^{2}}mN(N+1)(N+2) \cr
S_{F}&=& 
\frac{\alpha\rho}{2g^{2}}(N+1) tr\int \frac{d\Omega}{4\pi} 
(\frac{1}{\rho}\bar{\psi}\sigma^{i}L_{i}\psi 
+\bar{\psi}\sigma^{i}[a_{i},\psi] 
+\frac{2}{\rho}\bar{\psi}\psi)_{\star}, 
%\cr 
%&=& 
%\frac{\alpha\rho}{2g^{2}}(N+1) tr\int \frac{d\Omega}{4\pi}
%(\bar{\psi}D\psi+\bar{\psi}\sigma^{i}[a_{i},\psi] 
%+\frac{1}{\rho}\bar{\psi}\psi)_{\star} ,
\end{Eqnarray}
\noindent 
where $tr$ is taken over $m \times m$ matrices. 
%In the commutative limit, we obtain 
%\begin{equation}
%F_{ij}=-\frac{i}{\rho^{2}}K_{i}^{a}K_{j}^{b}
%(\partial_{a}b_{b}-\partial_{b}b_{a}+i[b_{a},b_{b}]).  
%\end{equation}

\vspace{0.5cm}

Let us consider a commutative limit. 
It is realized by the large $N$ limit with fixed $\rho$. 
The following relations are satisfied in the commutative limit: 
\begin{Eqnarray} 
\rho a_{i}(x)&=& K_{i}^{a}(x)b_{a}(x)
+\frac{x_{i}}{\rho}\phi  
 \label{killing}
\end{Eqnarray} 
where $i=1,2,3,a=\theta,\phi$ 
and $b_{a}$ is a gauge field on the sphere. 
Two fields $b_{a}$ are defined 
on the local coordinates $\theta,\phi$. 
$K_{i}^{a}$ are two Killing vectors and defined by 
\begin{Eqnarray} 
L_{i}&=& -iK_{i}^{a}(x)\partial_{a}.  
\end{Eqnarray} 
The explicit forms of these Killing vectors are given 
as follows 
\begin{equation}
\begin{array}{l l }
K_{1}^{\theta}=-\sin\phi  &
K_{1}^{\phi}=-\cot\theta\cos\phi \\ 
K_{2}^{\theta}=\cos\phi &
K_{2}^{\phi}=-\cot\theta\sin\phi  \\
K_{3}^{\theta}=0  &
K_{3}^{\phi}= 1  . \\
\end{array}
\end{equation}
From these Killing vectors, we obtain the metric tensor 
on $S^{2}$ as 
\begin{Eqnarray}
g^{ab}=K_{i}^{a}K_{i}^{b}. 
\end{Eqnarray}
Three fields 
$a_{i}$ are defined in three dimensional space {\bf R$^{3}$} 
and contain a gauge field $b_{a}$ on $S^{2}$ 
and a scalar field $\phi$ as well. 
In the commutative limit, we can separate the gauge field $b_{a}$ 
from $a_{i}$ using the Killing vectors as in (\ref{killing}). 
The bosonic part of the action is rewritten as 
\begin{Eqnarray}
S_{B}&=&-\frac{1}{4g_{YM}^{2}\rho^{2}}\int d\Omega(
K_{i}^{a}K_{j}^{b}K_{i}^{c}K_{j}^{d}
F_{ab}F_{cd}
+i2K_{i}^{a}K_{j}^{b}F_{ab}\epsilon_{ijk}
\frac{x_{k}}{\rho}\phi \cr 
&&\hspace{1.5cm}
+2K_{i}^{a}K_{i}^{b}
(D_{a}\phi)(D_{b}\phi)
-2\phi^{2}  )\cr 
&&-\frac{1}{2g_{YM}^{2}\rho^{2}}
\int d\Omega (i
\epsilon_{ijk}K_{i}^{a}K_{j}^{b}
F_{ab}\frac{x_{k}}{\rho}\phi
-\phi^{2}) \cr
%&=&-\frac{1}{2g_{YM}^{2}\rho^{2}}\int d\Omega(
%\frac{1}{\sin^{2}\theta}
%F_{\theta\phi}^{2}
%+\frac{4i}{\sin\theta}
%F_{\theta\phi}\phi 
%&&\hspace{1.5cm}
%+(D_{\theta}\phi)^{2}
%+\frac{1}{\sin^{2}\theta}(D_{\phi}\phi)^{2}
%-2\phi^{2} ) 
&=&-\frac{1}{2g_{YM}^{2}\rho^{2}}\int d^{2}x\sqrt{g}(
F_{ab}F^{ab}+\frac{2i}{\sqrt{g}}\epsilon^{ab}F_{ab}\phi
+(D_{a}\phi)(D^{a}\phi)-2\phi^{2})
\label{commutativeactionbosonic}
\end{Eqnarray}
where $F_{ab}=\frac{1}{i}\partial_{a}b_{b}
-\frac{1}{i}\partial_{b}b_{a}+[b_{a},b_{b}]$, 
$D_{a}=\frac{1}{i}\partial_{a}+[b_{a},\cdot]$ 
and $\epsilon^{\theta\phi}=-\epsilon^{\phi\theta}=1$. 
The fermionic part is also rewritten as 
\begin{Eqnarray}
S_{F}&=& 
\frac{1}{2g_{YM}^{2}\alpha^{3}\rho^{2}} \int d\Omega 
(\bar{\psi}\gamma^{a}D_{a}\psi 
 +\bar{\psi}\gamma_{3}[\phi,\psi] 
 +2\bar{\psi}\psi) 
\label{commutativeactionfermionic}
\end{Eqnarray}
where $\gamma_{3}=\sigma_{i}x_{i}/\rho$ (See (\ref{chiral}))
and $\gamma^{a}=\sigma^{i}K_{i}^{a}$. 
We thus obtained the action of 
a field theory with a gauge field, 
an adjoint scalar and a gaugino field on a sphere.

In the commutative limit, the gauge transformation 
becomes 
\begin{equation}
b_{a} \rightarrow b_{a} - \partial_{a}\lambda
\end{equation}
for $U(1)$ gauge group and 
\begin{equation}
b_{a} \rightarrow b_{a} - \partial_{a}\lambda
  +i[\lambda,b_{a}]
\end{equation}
for $U(m)$ gauge group. 
\vspace{0.6cm}

In the latter part of this section, we investigate the relation 
to a noncommutative gauge theory in a flat background 
by taking $\rho \rightarrow \infty$ limit
while fixing $\alpha$. 
By virtue of the $SO(3)$ symmetry, we may 
consider the theory around the north pole
without loss of generality. 
Around the north pole, 
$\hat{L}_{3}$ can be approximated as $\hat{L}_{3} \sim N/2$. 
By defining 
$\hat{L}_{i}^{\prime}=\sqrt{\frac{2}{N}}\hat{L}_{i}$, 
the commutation 
relation (\ref{su2}) becomes 
\begin{equation}
[\hat{L}_{1}^{\prime},\hat{L}_{2}^{\prime}] \sim i. \label{commflat}
\end{equation}

\noindent 
By further defining 
$\hat{x}_{i}^{\prime}= \alpha\hat{L}_{i}^{\prime}$ 
and $\hat{p}_{i}^{\prime}=\alpha^{-1} 
\varepsilon_{ij}\hat{L}_{j}^{\prime}$ 
$(i,j=1,2)$, 
we have 
\begin{equation} 
[\hat{x}_{1}^{\prime},\hat{x}_{2}^{\prime}]=i\alpha^{2}, 
\hspace{0.4cm}
[\hat{p}_{1}^{\prime},\hat{p}_{2}^{\prime}]=i\alpha^{-2}, 
\hspace{0.4cm}
[\hat{x}_{i}^{\prime},\hat{p}_{j}^{\prime}]=i\delta_{ij}. 
\end{equation}

\noindent 
We take the following limit to decompactify the sphere, 
\begin{Eqnarray}
\alpha =\mbox{fixed}, \hspace{0.4cm}  
\rho^{\prime} \gg 1 \hspace{0.2cm} (N \gg 1), \label{flat}
\end{Eqnarray}

\noindent 
where $\rho^{\prime2}=\hat{x}_{i}^{\prime}\hat{x}_{i}^{\prime}
=\frac{2}{N}\rho^{2}=\frac{N+2}{2}\alpha^{2}
\sim \frac{N}{2}\alpha^{2}$. 
In the coordinates of $\hat{x}_{i}^{\prime}$, 
the Plank constant is given by 
\begin{equation}
\frac{4\pi\rho^{\prime 2}}{N+1} \sim 2\pi\alpha^{2}.  
\end{equation}

\noindent 
$a \star b$ which is defined in (\ref{productmap}) 
becomes the Moyal product
$a \star_{M} b$  
because of (\ref{commflat}) and the Weyl type ordering property 
in (\ref{nonspheri}). 
The following replacements hold in this limit, 
\begin{Eqnarray}
\frac{1}{N+1}Tr \rightarrow \int \frac{d\Omega}{4\pi} 
= \int \frac{d^{2}x}{4\pi\rho^{2}} 
= \int \frac{d^{2}x^{\prime}}{4\pi\rho^{\prime2}} 
\end{Eqnarray}
\begin{equation}
Ad(\hat{p}_{i}^{\prime}) 
=\frac{1}{\rho^{\prime}}\varepsilon_{ij}Ad(\hat{L}_{j})
\rightarrow \frac{1}{i}\partial_{i}^{\prime}
\hspace{0.2cm} (i=1,2)
\end{equation}

\noindent 
We can regard $\hat{a}_{3}$ as the scalar field $\hat{\phi}$ 
around the north pole. 
For simplicity, only $U(1)$ case is treated in the 
present discussions. 
The bosonic part of the action (\ref{action}) becomes 
\begin{Eqnarray}
S_{B}&=&-\frac{\alpha^{4}}{2g^{2}}
Tr([\hat{L}_{1}+\rho\hat{a}_{1}, \hat{L}_{2}+\rho\hat{a}_{2}]^{2})
+\frac{\alpha^{4}}{2g^{2}} 
Tr(\rho\hat{\phi}[\hat{L}_{i}+\rho\hat{a}_{i},
[\hat{L}_{i}+\rho\hat{a}_{i},\rho\hat{\phi}]]) \cr
&&+\frac{2i\alpha^{4}}{g^{2}}
Tr([\hat{L}_{1}+\rho\hat{a}_{1},\hat{L}_{2}+\rho\hat{a}_{2}]\rho\hat{\phi}) \cr
&=&\frac{\alpha^{4}}{2g^{2}}(\rho^{\prime})^{4}
\{ 
-Tr([-\hat{p}_{2}^{\prime}-\hat{a}_{2}^{\prime}, 
\hat{p}_{1}^{\prime}+\hat{a}_{1}^{\prime}]^{2})
+Tr(\hat{\phi}^{\prime}
[\hat{p}_{i}^{\prime}+\hat{a}_{i}^{\prime},
[\hat{p}_{i}^{\prime}+\hat{a}_{i}^{\prime},\hat{\phi}^{\prime}]]) \cr
&&+\frac{4i}{\rho^{\prime}}
Tr ([-\hat{p}_{2}^{\prime}-\hat{a}_{2}^{\prime},
\hat{p}_{1}^{\prime}+\hat{a}_{1}^{\prime}]\hat{\phi}^{\prime}) 
\},
\end{Eqnarray}

\noindent 
where we have defined 
$\hat{a}_{i}=\sqrt{\frac{2}{N}}\hat{a}_{j}^{\prime}
\varepsilon_{ji} \hspace{0.1cm}(i,j=1,2)$ 
and $\hat{\phi}=\sqrt{\frac{2}{N}}\hat{\phi}^{\prime}$.
This action can be mapped to the following action, 
\begin{Eqnarray}
S_{B}&=&
\frac{\alpha^{6}N^{2}}{16\pi g^{2}}
\{ -\int d^{2}x^{\prime }F_{12}(x)\star F_{12}(x)
+\int d^{2}x^{\prime}\phi^{\prime}(x)(Ad(D))^{2}\phi^{\prime}(x) \cr
&&+\frac{4i}{\rho^{\prime}}
\int d^{2}x^{\prime} F_{12}(x)\star\phi^{\prime}(x) \} , \label{flatboson}
\end{Eqnarray}

\noindent 
where $Ad(D_{i})=[\frac{1}{i}\partial_{i}^{\prime}
+a_{i}^{\prime}, \cdot]$ and 
$F_{12}=\frac{1}{i} \partial_{1}^{\prime}a_{2}^{\prime}
-\frac{1}{i}\partial_{2}^{\prime}a_{1}^{\prime}
+[a_{1}^{\prime},a_{2}^{\prime}]_{\star}$. 
It is found that 
the Yang-Mills coupling is $g_{YM}^{2}=4\pi g^{2}/N^{2}\alpha^{6}$. 
Similarly the fermionic part of the action (\ref{action}) becomes
\begin{Eqnarray}
S_{F}&=&\frac{\alpha}{2g^{2}}Tr(\bar{\hat{\psi}}\sigma^{i}
  [\hat{L}_{i}+\rho\hat{a}_{i},\hat{\psi}] 
+2\bar{\hat{\psi}}\hat{\psi}) \cr 
&=&\frac{\alpha \rho^{\prime}}{2g^{2}}
Tr(\bar{\hat{\psi}}\tilde{\sigma}^{i}
  [\hat{p}_{i}^{\prime}+\hat{a}_{i}^{\prime},\hat{\psi}] 
+\bar{\hat{\psi}}\sigma^{3}
  [\hat{\phi}^{\prime},\hat{\psi}] 
+\frac{2}{\rho^{\prime}}\bar{\hat{\psi}}\hat{\psi})
\end{Eqnarray}

\noindent 
where we have defined $\tilde{\sigma}_{j}
=\varepsilon_{ji}\sigma_{i}\hspace{0.1cm}(i,j=1,2)$. 
This is also mapped as follows 
\begin{Eqnarray}
S_{F}
&=&\frac{\alpha \rho^{\prime}}{2 g^{2}}(N+1)
\int \frac{d^{2}x^{\prime}}{4\pi \rho^{\prime 2}} 
(\frac{1}{i}\bar{\psi}(x)
\tilde{\sigma}^{i}\partial_{i}^{\prime}\psi(x) \cr
&&+\bar{\psi}(x)
\tilde{\sigma}^{i}[a_{i}^{\prime}(x),\psi(x)]_{\star} 
+\bar{\psi}(x)\sigma^{3}[\phi^{\prime}(x),\psi(x)]_{\star}
+\frac{2}{\rho^{\prime}}\bar{\psi}(x)\psi(x)). \label{flatfermi}
\end{Eqnarray}

\noindent 
The gauge transformation (\ref{matrixgauge}) becomes
\begin{equation}
a_{i}^{\prime} \rightarrow 
a_{i}^{\prime}-\partial_{i}^{\prime}\lambda
  +i[\lambda,a_{i}^{\prime}]_{\star}. 
\end{equation}

\vspace{0.4cm}

We have seen that in the large radius 
and $\alpha$=fixed limit (\ref{flat}), 
the noncommutative Yang-Mills theory in the flat backgrounds 
can be obtained. 
The last term in (\ref{flatboson}) 
and the last term in (\ref{flatfermi}) 
become small  in the $\rho^{\prime} \rightarrow \infty$. 
To discuss the meaning of these terms, 
we integrate out the scalar field $\phi$. 
The action (\ref{flatboson}) becomes  
\begin{Eqnarray}
S&=&\frac{1}{4g_{YM}^{2}}
\{- \int d^{2}x^{\prime }F_{12}(x)\star F_{12}(x)
+\int d^{2}x^{\prime}\phi^{\prime}(x)(Ad(D))^{2}\phi^{\prime}(x) \cr
&&+\frac{4i}{\rho^{\prime}}
\int d^{2}x^{\prime} F_{12}(x)\star\phi^{\prime}(x) )\} \cr
&\rightarrow&
\frac{1}{4g_{YM}^{2}}
\{-\int d^{2}x^{\prime}F_{12}^{2}(x)
+\frac{4}{\rho^{\prime 2}}\int d^{2}x^{\prime}F_{12}(x)
(-\frac{1}{\partial^{2}})F_{12}(x) +O(a^{3})\} \cr 
&=&\frac{1}{4g_{YM}^{2}}
\{-\int d^{2}x^{\prime}F_{12}^{2}(x)
+\frac{4}{\rho^{\prime 2}}\int d^{2}x^{\prime}
d^{2}y^{\prime}F_{12}(x)
\Delta(x-y)F_{12}(y) +O(a^{3})\} \cr 
&=&\frac{1}{4 g_{YM}^{2}}\int d^{2}p
(-1+\frac{4}{\rho^{\prime 2}p^{2}})F_{12}(-p)F_{12}(p)
+O(a^{3}) , 
\end{Eqnarray}

\noindent 
where $-1/\partial^{2}=\Delta(x-y)
=\int d^{2}k e^{ik\cdot(x^{\prime}-y^{\prime})}/k^{2}$. 
The mass of the gauge field is found to be 
\begin{equation}
M_{\mbox{gauge}}=\frac{2}{\rho^{\prime}}. 
\end{equation}

\noindent 
We found that the gauge field 
has acquired the mass by absorbing the degree of freedom of the scalar field. 
%In the present situation,  the background restores supersymmetry 
%up to a $1/\rho^{\prime}$ correction. 
%
%In the $\rho^{\prime} \rightarrow \infty$ limit, 
%translation symmetry of $\psi$ is restored, 
%since Majorana spinor mass term drops. 
%By combining this symmetry and the original supersymmetry (\ref{susy}), 
%we can obtain ${\cal N}=2$ supersymmetry\cite{IKKT}. 
%This background preserve a half of this ${\cal N}=2$ supersymmetry. 
%
From (\ref{flatfermi}), the mass of the gaugino field is $2/\rho^{\prime}$  
which is degenerate with the mass of the gauge field. 
In the $\rho^{\prime} \rightarrow \infty$ limit, 
the gauge field and the gaugino field become massless. 

%%%%%%%%%%%%%%%%%%%%%%%%%%%%%%%%%%%%%%%%%%%%%%%%%%%%
%%%%%%%%%%%%%%%%%%%%%%%%%%%%%%%%%%%%%%%%%%%%%%%%%%%%

\section{Some Properties of Dirac Operator}
\hspace{0.4cm}

In this section, we investigate some properties of 
Dirac operator. 
Dirac operator in our model is given by 
\begin{equation}
\hat{D}=\frac{1}{\rho}(\sigma_{i}(Ad(\hat{L}_{i}) +1).
\end{equation}

\noindent 
Other Dirac operators are also constructed in \cite{WW,GM,GP}.
In the commutative limit, this operator becomes 
\begin{equation}
D=\frac{1}{\rho}(\sigma_{i}L_{i}+1), 
\end{equation}

\noindent 
which is the standard massless Dirac operator on the sphere\cite{jaye}. 
The second term of this operator comes from the contribution of 
the spin connection on the sphere. 
Around the north pole, the operator behaves as 
\begin{equation}
D=\sqrt{\frac{2}{N}}(
\tilde{\sigma}_{i}p_{i}^{\prime}+\frac{1}{\rho^{\prime}}),  
\end{equation}
and in the $\rho^{\prime} \rightarrow \infty$ limit, it approaches 
the massless Dirac operator in the flat background. 

We consider the following chirality operator 
\begin{equation}
\hat{\gamma}_{3}=\frac{1}{\rho}\sigma_{i}\hat{x}_{i}, 
\end{equation}
which becomes, in the commutative limit, 
\begin{equation}
\gamma_{3}=\frac{1}{\rho}\sigma_{i}x_{i}. \label{chiral}
\end{equation}
It is the standard chirality operator 
on the sphere\cite{jaye}. 
This chirality operator  obeys 
 $\hat{\gamma}_{3}^{2}=1 - 2 \hat{\gamma}_{3}/ \sqrt{N(N+2)}$ 
and does not satisfy the usual condition for the chirality operator
before taking the commutative limit. 
A chirality operator which  satisfies the condition $\hat{\gamma}_{3}^{2}=1$ 
and becomes (\ref{chiral}) in the commutative limit 
is given\cite{WW} by 
\begin{equation}
\hat{\gamma}_{3}^{\prime}
=\frac{1}{M}(\sigma_{i}\hat{x}_{i}+\frac{\alpha}{2}) 
\label{chiralwata} , 
\end{equation}

\noindent 
where $M=\alpha(N+1)/2$ is a normalization factor. 
Anticommutation relation between the Dirac operator and 
the chirality operator is 
\begin{equation}
\hat{D}\hat{\gamma}_{3}+\hat{\gamma}_{3}\hat{D}=\frac{2}{\rho^{2}}
\hat{x}_{i}Ad(\hat{L}_{i})
\end{equation}

\noindent 
which vanishes in the commutative limit 
as can be seen from (\ref{angope}). 
Another type of a Dirac operator which anticommutes with (\ref{chiralwata}) 
is constructed in \cite{WW}. 

We next investigate the spectrum of the Dirac operator.
We consider an eigenstate $\Psi$ of $J^{2}$ and $J_{3}$, 
satisfying $J^{2}\Psi_{jm}=j(j+1)\Psi_{jm}$ 
and $J_{3}\Psi_{jm}=(m+1/2)\Psi_{jm}$, where  
$J_{i}=L_{i}+\sigma_{i}/2$ and $1/2 \le j \le N+1/2$ and 
$-l\le m \le l$. 
Acting on $\hat{\Psi}_{jm}$, 
square of the Dirac operator becomes 
\begin{Eqnarray}
\rho^{2}\hat{D}^{2}\hat{\Psi}_{jm}
&=& \sigma_{i}[\hat{L}_{i}, \hat{\Psi}_{jm}] 
+[\hat{L}_{i},[\hat{L}_{i},\hat{\Psi}_{jm}]] +\hat{\Psi}_{jm} \cr
&=& ((Ad(J))^{2}-(Ad(L))^{2}-\frac{3}{4})\hat{\Psi}_{jm} 
+(Ad(L))^{2}\hat{\Psi}_{jm}+ \hat{\Psi}_{jm} \cr 
&=& (j+\frac{1}{2})^{2} \hat{\Psi}_{jm}.
\end{Eqnarray}

\noindent 
We have used the relation 
$J^{2}=L^{2}+L\cdot \sigma + 3/4$.  
This spectrum is identical with the spectrum in the commutative limit. 

%%%%%%%%%%%%%%%%%%%%%%%%%%%%%%%%%%%%%%%%%%%%%%%%%%%%
%%%%%%%%%%%%%%%%%%%%%%%%%%%%%%%%%%%%%%%%%%%%%%%%%%%%

\section{Stability of Classical Solutions}
\hspace{0.4cm}
We have argued the noncommutative 
gauge theories on the fuzzy sphere. 
This section is devoted to 
a discussion of stability of the classical solutions. 
Let us first evaluate classical values of the action  
for the solutions of  
(\ref{diagsol}) and (\ref{clsol}). 

The action vanishes 
\begin{equation}
S=0, 
\end{equation}

\noindent 
for commuting matrices (\ref{diagsol}), and becomes 
\begin{Eqnarray}
S&=&-\frac{1}{24g^{2}}N(N+1)(N+2)\alpha^{4} \cr 
&=&-\frac{2}{3g^{2}}\rho^{4}\frac{N+1}{N(N+2)},  \label{classvalue} 
\end{Eqnarray}

\noindent 
for the fuzzy sphere
\footnote{The classical value of the action depends on 
the quadratic Casimir. 
That is, it depends on the representation.
We can construct an alternative representation in (\ref{rep}). 
The value of the action is larger than the irreducible representation. 
Therefore the solutions of reducible representation 
are unstable\cite{myers}.} (\ref{clsol}).
The solution of the fuzzy sphere has lower energy than 
the solution of commuting matrices, 
and therefore the fuzzy sphere is more stable than the commuting matrices 
at the classical level. 

We next investigate one-loop effective action\cite{IKKT,AIKKT} 
around the classical solutions. 
We decompose the matrices $A_{i}$ and $\psi$ into the classical backgrounds 
and fluctuations: 
\begin{Eqnarray}
A_{i}&=&X_{i}+\tilde{A}_{i}, \cr 
\psi&=&\chi +\tilde{\psi}. 
\end{Eqnarray}

\noindent 
We add the following gauge-fixing term\cite{IKKT}, 
\begin{Eqnarray}
&&S_{g.f.}=-\frac{1}{g^{2}}Tr(\frac{1}{2}[X_{i},A_{i}]^{2}+[X_{i},b][A_{i},c]), 
\end{Eqnarray}
\noindent 
where $b$ and $c$ are ghost and anti-ghost, respectively. 
We expand the action up to the second order of the fluctuations 
and set $\chi=0$:  
\begin{Eqnarray} 
S&=&-\frac{1}{g^{2}}Tr(\frac{1}{2}[X_{i},\tilde{A}_{j}]^{2}+
 [X_{i},X_{j}][\tilde{A}_{i},\tilde{A}_{j}] \cr 
&&-i\alpha\epsilon_{ijk}X_{i}[\tilde{A}_{j},\tilde{A}_{k}] \cr
&&-\frac{1}{2}\bar{\tilde{\psi}}\Gamma^{i}[X_{i},\tilde{\psi}]
-\alpha\bar{\tilde{\psi}}\tilde{\psi} +[X_{i},b][X_{i},c] ). 
\end{Eqnarray} 

\noindent 
Then the one-loop effective action is calculated as 
\begin{equation}
W=-\log \int d\tilde{A} d\tilde{\psi} dbdc e^{-S}. 
\end{equation}
We first consider the one-loop effective action for the 
commuting matrices.
The contributions from $\tilde{A}$ and $\tilde{\psi}$ 
are calculated as follows 
\begin{Eqnarray}
W_{B}&=&\frac{1}{2}\sum_{i \neq j}(\log(x^{(i)}-x^{(j)})^{2} 
+\log(1-\frac{4\alpha^{2}}{(x^{(i)}-x^{(j)})^{2} })),  \cr 
W_{F}&=&-\frac{1}{2}\sum_{i \neq j}(\log(x^{(i)}-x^{(j)})^{2} 
+\log(1-\frac{4\alpha^{2}}{(x^{(i)}-x^{(j)})^{2} })). 
\end{Eqnarray}
\noindent  
Contributions from ghosts are included in $W_{B}$. 
We find that the one-loop effective action for commuting matrices 
vanishes due to the cancellation between bosonic and fermionic contributions 
as expected from supersymmetry. 
Therefore there are no net forces between the eigenvalues 
if we neglect the effect of the fermionic zero modes\cite{AIKKT}. 
For the fuzzy sphere, 
one-loop effective action can be calculated as  
\begin{Eqnarray}
W_{B}&=& \frac{1}{2}Tr\log (Ad(L))^{2}, \cr 
W_{F}&=& -\frac{1}{2}Tr\log (Ad(L))^{2} 
-\frac{1}{4}Tr\log (1+\frac{4+3\sigma^{i}Ad(L_{i})}{(Ad(L))^{2}}). 
\end{Eqnarray}

\noindent 
In this case, the one-loop effective action does not vanish because 
this background preserves no supersymmetry. 
One-loop quantum corrections to the classical action 
(\ref{classvalue}) are 
\begin{equation}
W=-\frac{1}{4}Tr\log (1+\frac{4+3\sigma^{i}Ad(L_{i})}{(Ad(L))^{2}}).
\end{equation}

\noindent 
Let us evaluate this quantity. 
This expression can be rewritten using a replacement 
$\sigma^{i}Ad(L_{i}) \rightarrow j(j+1)-l(l+1) -3/4=s(2l+1)-1/2$, where $s=\pm 1/2$: 
\begin{Eqnarray}
W&=&-\frac{1}{4}\sum_{l}^{N}\sum_{s=\pm 1/2}(2l+1)
\log (1+\frac{3s(2l+1)+\frac{5}{2}}{l(l+1)}) \cr 
&=& -\frac{1}{2}\sum_{l}^{N}(2l+1)
\log \frac{(l+2)(l-1)}{l(l+1)} .
\end{Eqnarray}

\noindent 
%For $1/l\ll 1$, a perturbative expansion is applied 
%to evaluate $W$. 
In the large $N$ limit, 
%small $l$ constibution can be neglected and 
we have
\begin{Eqnarray}
W
%\frac{1}{4}\sum_{l \gg 1}^{N}\frac{34}{l} 
 \sim 2 \log N .
\end{Eqnarray}
\noindent 
The value of the effective action including 
both the classical and the one-loop quantum corrections 
becomes 
\begin{Eqnarray}
S_{eff}&=&-\frac{2}{3g^{2}}\rho^{4}\frac{N+1}{N(N+2)} 
+2 \log N \cr 
&=& -\frac{\pi}{6\rho^{2}g_{YM}^{2}}N(N+2) 
+2 \log N. 
\end{Eqnarray}
\noindent 
For large enough $N$  and  $g_{YM}$=fixed, 
the fuzzy sphere is 
more stable than the commuting matrices
\footnote{
As studied in papers\cite{Staudacher},
the partition function of $d=3$ supersymmetric reduced models is not
convergent against integral over bosonic variables.
It implies that eigenvalues are more likely to be separated from 
each other and the fuzzy sphere solution is unstable nonperturbatively,
in contrast to our perturbative analysis.
From this point of view, it is more interesting if we can construct
nontrivial curved backgrounds from convergent higher dimensional
supersymmetric reduced models. 
We thank Staudacher for informing us of their analysis on  
reduced model integrals. }. 
On the other hand, 
there is a region of $S_{eff} > 0$ 
if we take $g_{YM}$  sufficiently large so that the second term 
dominates the first one. 
In the region, the commuting matrices is more stable 
than the fuzzy sphere. 

%%%%%%%%%%%%%%%%%%%%%%%%%%%%%%%%%%%%%%%%%%%%%%%%%%%%%%%%%%%%%%%%%%
%%%%%%%%%%%%%%%%%%%%%%%%%%%%%%%%%%%%%%%%%%%%%%%%%%%%%%%%%%%%%%%%%%
%%%%%%%%%%%%%%%%%%%%%%%%%%%%%%%%%%%%%%%%%%%%%%%%%%%%%%%%%%%%%%%%%%

\section{Supersymmetric Noncommutative Gauge Theory \\
on Fuzzy Sphere}
\hspace{0.4cm}
In this section, we consider another action 
which gives a supersymmetric noncommutative gauge theory 
on the fuzzy sphere. 
Supersymmetric gauge theories on a fuzzy sphere is 
also considered in \cite{9903112}. 
The action is defined by 

\begin{equation}
 S= \frac{1}{g^{2}} Tr ( -\frac{1}{4} [ A_{i} ,A_{j}] 
         [A^{i} ,A^{j} ]  
  +\frac{2}{3}i\alpha \epsilon_{ijk}A^{i}A^{j}A^{k} 
  +\frac{1}{2}\bar{\psi}\sigma^{i}[A_{i},\psi] ) . \label{modaction}
\end{equation}
%Other symmetries (\ref{translation}) and (\ref{gaugetr}) are 
%also preserved in this model. 
It has has the following ${\cal N}=2$ supersymmetry: 

\begin{Eqnarray}
&&\delta^{(1)} A_{i} =i \bar{\epsilon}\sigma_{i}\psi \cr
&&\delta^{(1)} \psi  =\frac{i}{2}
([A_{i} ,A_{j} ] -i\alpha\epsilon_{ijk}A_{k})
\sigma^{ij}\epsilon , \label{susy21}
\end{Eqnarray}
and 
\begin{Eqnarray}
&&\delta A_{i}^{(2)} =0 \cr
&&\delta ^{(2)}\psi  =\xi .  \label{susy22}
\end{Eqnarray}
The transformation for the gaugino in (\ref{susy21})
is modified in comparison to the transformation 
in the usual Yang-Mills reduced models
\footnote{ 
In \cite{hashi}, there is a discussion about 
the supersymmetry transformation on the fuzzy sphere. 
In the context of string theory, the spherical D2-brane can 
be a supersymmetric configuration. 
The fact that our model (\ref{modaction}) 
has the modified supersymmetry (\ref{susy21}) 
which vanishes for the fuzzy sphere 
is consistent with the comment given in \cite{hashi}.
We thank K.Hashimoto and K.Krasnov for their stimulating 
comments. }. 
Let us check that these two transformations form 
the ${\cal N}=2$ supersymmetry algebra. 
Algebra is also modified due to the modification 
of supersymmetry. 
We have the following relations:

\begin{Eqnarray}
(\delta_{\epsilon^{1}}^{(1)}\delta_{\epsilon^{2}}^{(1)}
-\delta_{\epsilon^{2}}^{(1)}\delta_{\epsilon^{1}}^{(1)})
\psi&=& i[\psi,\lambda]+i\theta_{i}\frac{\sigma_{i}}{2}\psi , \cr 
(\delta_{\epsilon^{1}}^{(1)}\delta_{\epsilon^{2}}^{(1)}
-\delta_{\epsilon^{2}}^{(1)}\delta_{\epsilon^{1}}^{(1)})
A_{i}&=& i[A_{i},\lambda]+\epsilon_{ijk}\theta_{j}A_{k} , \label{newalgebra}
\end{Eqnarray}
where $\lambda=2i(\bar{\epsilon}_{2}\sigma_{j}\epsilon_{1})A_{j}$ 
and $\theta_{j}=2i\alpha(\bar{\epsilon}_{2}\sigma_{j}\epsilon_{1})$. 
The second term in the right hand side is a new term corresponding to
$SO(3)$ rotation. 
Other commutation relations are calculated as follows, 
\begin{Eqnarray}
(\delta_{\epsilon}^{(1)}\delta_{\xi}^{(2)}
-\delta_{\xi}^{(2)}\delta_{\epsilon}^{(1)})
\psi&=&0, \cr 
(\delta_{\epsilon}^{(1)}\delta_{\xi}^{(2)}
-\delta_{\xi}^{(2)}\delta_{\epsilon}^{(1)})
A_{i}&=& -i\bar{\epsilon}\sigma_{i}\xi ,
\end{Eqnarray}
and 
\begin{Eqnarray}
(\delta_{\xi^{1}}^{(2)}\delta_{\xi^{2}}^{(2)}
-\delta_{\xi^{2}}^{(2)}\delta_{\xi^{1}}^{(2)})
\psi&=&0 , \cr 
(\delta_{\xi^{1}}^{(2)}\delta_{\xi^{2}}^{(2)}
-\delta_{\xi^{2}}^{(2)}\delta_{\xi^{1}}^{(2)})
A_{i}&=&0 .
\end{Eqnarray}
If we take a linear combination of $\delta^{(1)}$ 
and $\delta^{(2)}$ as 

\begin{Eqnarray}
\tilde{\delta}^{(1)}&=&
\delta^{(1)}+\delta^{(2)} , \cr
\tilde{\delta}^{(2)}&=& 
i(\delta^{(1)}-\delta^{(2)}),  
\end{Eqnarray}
we can obtain the following commutation relations 
up to a gauge symmetry and $SO(3)$ symmetry, 
\begin{Eqnarray}
(\tilde{\delta}_{\epsilon}^{(1)}\tilde{\delta}_{\xi}^{(1)}
-\tilde{\delta}_{\xi}^{(1)}\tilde{\delta}_{\epsilon}^{(1)})
\psi&=&
%i\theta_{i}\frac{\sigma_{i}}{2}\psi
0, \cr 
(\tilde{\delta}_{\epsilon}^{(1)}\tilde{\delta}_{\xi}^{(1)}
-\tilde{\delta}_{\xi}^{(1)}\tilde{\delta}_{\epsilon}^{(1)})
A_{i}&=& -2i\bar{\epsilon}\sigma_{i}\xi 
%+\epsilon_{ijk}\theta_{j}A_{k} 
 , \cr
(\tilde{\delta}_{\epsilon}^{(2)}\tilde{\delta}_{\xi}^{(2)}
-\tilde{\delta}_{\xi}^{(2)}\tilde{\delta}_{\epsilon}^{(2)})
\psi &=&
%-i\theta_{i}\frac{\sigma_{i}}{2}\psi
 0, \cr 
(\tilde{\delta}_{\epsilon}^{(2)}\tilde{\delta}_{\xi}^{(2)}
-\tilde{\delta}_{\xi}^{(2)}\tilde{\delta}_{\epsilon}^{(2)})
A_{i}&=&-2i\bar{\epsilon}\sigma_{i}\xi 
%-\epsilon_{ijk}\theta_{j}A_{k}
 ,\cr
(\tilde{\delta}_{\epsilon}^{(1)}\tilde{\delta}_{\xi}^{(2)}
-\tilde{\delta}_{\xi}^{(2)}\tilde{\delta}_{\epsilon}^{(1)})
\psi&=&
%-\theta_{i}\frac{\sigma_{i}}{2}\psi 
0, \cr 
(\tilde{\delta}_{\epsilon}^{(1)}\tilde{\delta}_{\xi}^{(2)}
-\tilde{\delta}_{\xi}^{(2)}\tilde{\delta}_{\epsilon}^{(1)})
A_{i}&=&
%i\epsilon_{ijk}\theta_{j}A_{k} 
0.
\end{Eqnarray}
%where $\theta_{i}=\bar{\xi}\sigma_{i}\epsilon$. 
We find that these commutation relations indeed show the 
${\cal N}=2$ supersymmetry algebra. 
A new feature is the appearance of $SO(3)$ rotation. 
% in the right hand side.
This model has the same classical solutions as the previous model, 
commuting diagonal matrices and the fuzzy sphere. 
In this model, the fuzzy sphere preserves 
half of the 
${\cal N}=2$ supersymmetries 
since (\ref{susy21}) vanishes for the fuzzy sphere while 
(\ref{susy22}) does not, and 
this solution corresponds 
to a $1/2$ BPS background. 
Looking at the algebra (\ref{newalgebra}), 
the remaining supersymmetry on the fuzzy sphere 
generates $SO(3)$ rotation instead of 
a constant shift of $A_{i}$. 
It is natural since translation on a sphere is 
generated by $SO(3)$ rotation. 
On the other hand, commuting matrices break
all the supersymmetry. 

By expanding the bosonic matrices around 
the fuzzy sphere solution (\ref{clsol}) 
as in (\ref{bosonicmatexpansion}) 
and applying the mapping rule which is given in section 2, 
we can obtain an ${\cal N}=1$ 
supersymmetric $U(1)$ or $U(n)$ noncommutative 
gauge theory on the fuzzy sphere. 
For simplicity only $U(1)$ 
gauge theory is considered in the following. 
The bosonic part of the action is 
\begin{Eqnarray}
S_{B}&=&-\frac{\alpha^{4}\rho^{4}}{4g^{2}}
Tr(\hat{F}_{ij}\hat{F}_{ij}) \cr 
&&-\frac{i}{2g^{2}}\alpha^{4}\rho^{3}\epsilon^{ijk}
Tr(\frac{1}{\rho}[\hat{L}_{i},\hat{a}_{j}]\hat{a}_{k}
+\frac{1}{3}\hat{a}_{i}[\hat{a}_{j},\hat{a}_{k}] 
-\frac{i}{2\rho}\epsilon_{ijm}\hat{a}^{m}\hat{a}_{k})
-\frac{\alpha^{2}}{6g^{2}}Tr({\hat{x}^{i}\hat{x}_{i}}) \cr
&=&
-\frac{\alpha^{4}\rho^{4}}{4g^{2}}(N+1)\int \frac{d\Omega}{4\pi}
(F_{ij}F_{ij})_{\star} \cr 
&&-\frac{i}{2g^{2}}\alpha^{4}\rho^{3}\epsilon^{ijk}
(N+1)\int \frac{d\Omega}{4\pi}
(\frac{1}{\rho}(L_{i}a_{j})a_{k}
+\frac{1}{3}a_{i}[a_{j},a_{k}] 
-\frac{i}{2\rho}\epsilon_{ijm}a^{m}a_{k})_{\star} \cr 
&&-\frac{\alpha^{4}}{24g^{2}}N(N+1)(N+2), 
\end{Eqnarray}
where 
$F_{ij}$ is defined in (\ref{Fieldstrength}). 

The fermionic part is 
\begin{Eqnarray}
S_{F}&=&\frac{\alpha}{2g^{2}}Tr\bar{\hat{\psi}}\sigma^{i}
  [\hat{L}_{i}+\rho\hat{a}_{i},\hat{\psi}]  \cr 
&=& 
\frac{\alpha\rho}{2g^{2}}(N+1)
\int \frac{d\Omega}{4\pi} (
\frac{1}{\rho}\bar{\psi}\sigma^{i}L_{i}\psi 
+\bar{\psi}\sigma^{i}[a_{i},\psi] )_{\star}. 
\end{Eqnarray}
%where $D=\frac{1}{\rho}(\sigma^{i}L_{i}+1)$ is a massless 
%Dirac operator on a sphere. 
%The gaugino field  has a mass term 
%on the sphere due to 
%the contribution from the spin connection. 

This supersymmetric noncommutative gauge theory 
has the following ${\cal N}=1$ supersymmetry: 
\begin{Eqnarray} 
&&\delta a_{i} =\frac{i}{\alpha\rho}\bar{\epsilon}\sigma_{i}\psi \cr 
&&\delta \psi  =\frac{i\alpha^{2}\rho^{2}}{2}F_{ij} 
\sigma^{ij}\epsilon . 
\label{susyn=1}
\end{Eqnarray} 
Applying this transformation twice generates 
translation on the sphere, 
that is the rotation. 

Let us consider a commutative limit. 
The action in the commutative limit is obtained 
by the same procedure as in section 2. 
In terms of the gauge field $b_{a}$ 
and the scalar field $\phi$, 
the bosonic part becomes 
\begin{Eqnarray}
S_{B}=-\frac{1}{2g_{YM}^{2}\rho^{2}}\int d^{2}x\sqrt{g}(
F_{ab}F^{ab}+\frac{2i}{\sqrt{g}}\epsilon^{ab}F_{ab}\phi
+(D_{a}\phi)(D^{a}\phi)-2\phi^{2}).
\end{Eqnarray}
This is  the same as 
(\ref{commutativeactionbosonic}). 
The fermionic part becomes 
\begin{Eqnarray}
S_{F}&=& 
\frac{1}{2g_{YM}^{2}\rho^{2}} \int d\Omega 
(\bar{\psi}\gamma^{a}D_{a}\psi 
 +\bar{\psi}\gamma_{3}[\phi,\psi] ).  
\end{Eqnarray}
We have rescaled as $\psi \rightarrow \alpha^{3/2}\psi$. 
The difference between 
the action obtained in this section and 
the action obtained in section 2 is that 
the former is supersymmetric while the latter is not. 
The ${\cal N}=1$ supersymmetry transformation 
(\ref{susyn=1}) is rewritten 
in terms of the gauge field and the scalar field 
as 
\begin{Eqnarray} 
&&\delta \phi =i\bar{\epsilon^{\prime}}
\gamma_{3}\psi \cr 
&&\delta b^{a}=i\bar{\epsilon^{\prime}}
\gamma^{a}\psi 
\end{Eqnarray} 
and 
\begin{Eqnarray} 
\delta \psi =
\frac{i}{2}
(F_{ab}\gamma^{ab}
-2\rho\gamma^{3}\phi
+\rho (D_{a}\phi)[\gamma^{3},\gamma^{a}]
)\epsilon^{\prime} 
\end{Eqnarray} 
where $b^{a} \equiv g^{ab}b_{b}$, 
$\gamma^{3}=x_{i}\sigma_{i}/\rho$, 
$\gamma^{a}=K_{i}^{a}\sigma_{i}$ and 
$\epsilon^{\prime} =\alpha^{1/2}\epsilon$. 

We next consider stability of the two classical solutions 
against quantum corrections in the same manner as discussed in section 4. 
One-loop corrections to the commuting matrices are calculated as 

\begin{Eqnarray} 
W_{B}&=&\frac{1}{2}\sum_{i \neq j}(\log(x^{(i)}-x^{(j)})^{2}  
+\log(1-\frac{4\alpha^{2}}{(x^{(i)}-x^{(j)})^{2} })),  \cr  
W_{F}&=&-\frac{1}{2}\sum_{i \neq j}\log(x^{(i)}-x^{(j)})^{2} . 
\end{Eqnarray}
Then, the value of the effective action including both 
classical and one-loop quantum effects is given by 
\begin{Eqnarray}
S_{eff}^{CM}=\frac{1}{2}\sum_{i \neq j}
\log(1-\frac{4\alpha^{2}}{(x^{(i)}-x^{(j)})^{2} }). \label{effCM}
\end{Eqnarray}
From this expression, we can read off that 
the eigenvalues tend to collapse into a single 
ball whose size is of order $\alpha$.  
This shows that there exist a minimum length scale 
set by $\alpha$ in this theory. 
Since there are $N(N+1)/2$ pairs of eigenvalues, 
$S_{eff}^{CM}$ is negative and of order $N^{2}$.

For the fuzzy sphere, one-loop correction can be calculated as 
\begin{Eqnarray} 
W_{B}&=& \frac{1}{2}Tr\log (Ad(L))^{2}, \cr 
W_{F}&=& -\frac{1}{2}Tr\log (Ad(L))^{2}, 
\end{Eqnarray}
and the effective action for 
the fuzzy sphere does not receive quantum corrections 
perturbatively as expected. 
Therefore the value of the effective action is 
\begin{Eqnarray}
S_{eff}^{FS}=-\frac{2}{3g^{2}}\rho^{4}\frac{N+1}{N(N+2)}  
= -\frac{\pi}{6\rho^{2}g_{YM}^{2}}N(N+2) .  \label{effFS}
\end{Eqnarray}
In this case, since both of (\ref{effCM}) and (\ref{effFS}) 
are negative valued and of the same order $N^{2}$, 
it is difficult to conclude perturbative stability of the 
fuzzy sphere at this level. 

%%%%%%%%%%%%%%%%%%%%%%%%%%%%%%%%%%%%%%%%%%%%%%%%%%%%%%%%%%%%%%%%%%%%%
%%%%%%%%%%%%%%%%%%%%%%%%%%%%%%%%%%%%%%%%%%%%%%%%%%%%%%%%%%%%%%%%%%%%%

\section{Summary}
\hspace{0.4cm}
In this paper, we have studied  noncommutative gauge theories 
on the fuzzy sphere. 
We considered two different types of supersymmetric
three-dimensional matrix model
actions with a Chern-Simons term. 
These models have a classical solution which represents a fuzzy sphere. 
By expanding the models around this solution, we have obtained 
noncommutative gauge theories on the fuzzy sphere. 
The gauge field acquires a mass due to the Chern-Simons term.
%The mass is degenerate with the gaugino mass. 

We have discussed two large $N$ limits. 
One corresponds to a commutative limit. 
By taking this limit, we obtained a gauge theory 
on a commutative sphere. 
Another limit corresponds to a decompactifying limit. 
A noncommutative gauge theory on the fuzzy sphere 
becomes a noncommutative gauge theory 
on a noncommutative plane. 

Two types of three-dimensional supersymmetric matrix model actions 
we have considered in this paper have
different supersymmetry properties.
The first type contains  a Majorana mass term in order to preserve
the original supersymmetry in the flat space.
The second one does not have any other term than the Yang Mills and
the Chern Simons terms but it is invariant under a modified 
supersymmetry. 
The fuzzy sphere preserves  supersymmetry in the second type 
while it does not in the first one. 
Therefore the second one will be  more natural from the fuzzy 
sphere point of view.
On the other hand, 
as for the commuting matrices, the situation is opposite. 
A solution with commuting matrices  preserves the supersymmetry
in the first type and eigenvalues can 
move freely at least perturbatively.
In the second type, 
eigenvalues collapse into a small ball whose size is of 
the noncommutative scale. 
\par
We have also investigated the stability of the fuzzy sphere
against quantum fluctuations. By calculating the one-loop 
effective action, we showed that in the first model 
the fuzzy sphere is stable for fixed $g_{YM}$ in the large $N$ limit. 
However, if we take $g_{YM}$ sufficiently large with the 
large $N$ limit, the fuzzy sphere becomes unstable and 
decays into a set of diagonal eigenvalues. 
On the other hand, the fuzzy sphere is stable 
against quantum corrections because of the nature of a BPS state
in the second model. 

Finally let us comment on ambiguities of  operator orderings
and corresponding freedom for mapping from matrices to functions.  
Matrices on the fuzzy sphere 
are expanded by a noncommutative spherical harmonics
as discussed in section 2.
The ordering of the noncommutative coordinates 
in the noncommutative spherical harmonics 
corresponds to the Weyl type ordering. 
Therefore a product of the noncommutative spherical harmonics 
becomes Moyal product in the decompactifying limit. 
On the other hand,
a stereographic projection from a fuzzy sphere 
to a noncommutative complex plane as discussed in the appendix B 
enables us to construct a normal ordered type basis. 
After mapping from matrices to functions, 
a product of  functions are written by the so called Berezin product.
Since such noncommutative products are known for general K\"{a}hler
manifolds, 
it would be an interesting problem to 
obtain noncommutative gauge theories on more general curved backgrounds
from  matrix models.
%%%%%%%%%%%%%%%%%%%%%%%%%%%%%%%%%%%%%%%%%%%%%

%%%%%%%%%%%%%%%%%%%%%%%%%%%%%%%%%%%%%%%%%%%%%

%%%%%%%%%%%%%%%%%%%%%%%%%%%%%%%%%%%%%%%%%
\renewcommand{\theequation}{\Alph{section}.\arabic{equation}}
%%%%%%%%%%%%%%%%%%%%%%%%%%%%%%%%%%%%%%%%%%

%%%%%%%%%%%%%%%%%%%%%%%%%%%%%%%%%%%%%%%%%%%%%
\appendix
\section{Star product on fuzzy sphere}
\setcounter{equation}{0}
\label{sec:starpro}
\hspace{0.4cm}
In this section we summarize the noncommutative products on fuzzy sphere. 
We mainly follow \cite{Hoppe}. 
We have expanded functions and matrices as (\ref{function}) 
and (\ref{matrix}):

\begin{equation}
a(\Omega)=\sum_{l=0}^{l=N}\sum_{m=-l}^{l}
a_{lm}Y_{lm}(\Omega), \label{functionA} 
\end{equation}

\begin{equation}
\hat{a}=\sum_{l=0}^{l=N}\sum_{m=-l}^{l}
a_{lm}\hat{Y}_{lm}. \label{matricesB}
\end{equation}

%\begin{equation}
%a_{lm}=\int \Omega Y_{lm}^{\ast}a(\Omega)
%\end{equation}

%\begin{equation}
%a_{lm}=\frac{1}{N+1}Tr(\hat{Y}_{lm}^{\dagger}\hat{a})
%\end{equation}

\noindent 
Normalization is fixed as in (\ref{nor1}) and (\ref{normali}). 
Combining (\ref{functionA}) and (\ref{matricesB}), we can obtain a map between 
the field and the function:

\begin{equation}
a(\Omega)=\frac{1}{N+1}\sum_{lm}
Tr(\hat{Y}_{lm}^{\dagger}\hat{a})Y_{lm}(\Omega).
\end{equation}

\noindent 
Let us consider the product of the two spherical harmonics, 
$\hat{Y}_{lm}$ and $\hat{Y}_{l^{\prime}m^{\prime}}$. 
We have required that maximum value of $l$ is $N$. 
This product is expanded by the spherical harmonics and it contains 
$\hat{Y}_{l+l^{\prime}}$. 
We assume that $N$ is large such that 
$l+l^{\prime}$ does not exceed $N$.  
We define the noncommutative product on the fuzzy sphere as 

\begin{Eqnarray}
a\star b(\Omega)&\equiv&\frac{1}{N+1}\sum_{lm} 
Tr(\hat{a}\hat{b}\hat{Y}_{lm}^{\dagger})Y_{lm}(\Omega) \cr 
&=&\frac{1}{N+1}\sum_{lm}\sum_{l^{\prime}m^{\prime}}
\sum_{l^{\prime\prime}m^{\prime\prime}} 
\int d\Omega^{\prime}d\Omega^{\prime\prime}
Y_{l^{\prime}m^{\prime}}^{\ast}(\Omega^{\prime})
Y_{l^{\prime\prime}m^{\prime\prime}}^{\ast}(\Omega^{\prime\prime})
a(\Omega^{\prime})b(\Omega^{\prime\prime}) \cr
&&Tr(\hat{Y}_{l^{\prime}m^{\prime}}
\hat{Y}_{l^{\prime\prime}m^{\prime\prime}}
\hat{Y}_{lm}^{\dagger})Y_{lm}(\Omega).  \label{product}
\end{Eqnarray}

\noindent 
Components of the spherical harmonics can be given using Wigner Eckart theorem:

\begin{equation}
(\hat{Y}_{lm})_{ss^{\prime}} 
=<\frac{N}{2},s |\hat{Y}_{lm}|\frac{N}{2},s^{\prime}>
=(-1)^{\frac{N}{2}-s}
 \left( \begin{array}{c c c}
  \frac{N}{2} & l & \frac{N}{2} \\ -s  & m & s^{\prime} \\
 \end{array} \right)
\sqrt{(2l+1)(N+1)}
\end{equation}

\noindent 
where $s,s^{\prime}=-N/2,-N/2+1, \cdots, 0, \cdots, N/2$. 

\noindent 
Then we can calculate the trace of three spherical harmonics: 

\begin{Eqnarray}
&&\frac{1}{N+1}Tr(\hat{Y}_{l_{1}m_{1}}\hat{Y}_{l_{2}m_{2}}
\hat{Y}_{l_{3}m_{3}}) \cr 
&=&(-1)^{N+l_{1}+l_{2}+l_{3}}
\left( \begin{array}{c c c}
  l_{1} & l_{2} & l_{3} \\ m_{1}  & m_{2} & m_{3} \\
 \end{array} \right)
\left\{ \begin{array}{c c c}
  l_{1} & l_{2} & l_{3} \\ \frac{N}{2}  & \frac{N}{2} & \frac{N}{2} \\
 \end{array} \right\}
\sqrt{(2l_{1}+1)(2l_{2}+1)(2l_{3}+1)(N+1)}.
\end{Eqnarray}

\noindent 
where $(\cdots)$ and $\{\cdots\}$ are Wigner's 3$j$-symbol and 
6$j$-symbol respectively. 
$\{\cdots\}$ behaves as $N^{-3/2}$ for $N \rightarrow \infty$\cite{Hoppe}.
$(\cdots)$ becomes zero only when $m_{1}+m_{2}+m_{3}=0$.
By substituting this quantity into (\ref{product}), 
we get the explicit expression of the noncommutative product 
on the fuzzy sphere. 

Commutator of two matrices is 

\begin{Eqnarray}
[\hat{a},\hat{b}]&=&\sum_{l_{1}m_{1}}\sum_{l_{2}m_{2}}
a_{l_{1}m_{1}}b_{l_{2}m_{2}}
[\hat{Y}_{l_{1}m_{1}},\hat{Y}_{l_{2}m_{2}}] \cr 
&=&\sum_{l_{1}m_{1}}\sum_{l_{2}m_{2}}\sum_{l_{3}m_{3}}
a_{l_{1}m_{1}}b_{l_{2}m_{2}}
f_{l_{1}m_{1}l_{2}m_{2}}^{l_{3}m_{3}}
\hat{Y}_{l_{3}m_{3}}. 
\end{Eqnarray}

\noindent 
$f_{l_{1}m_{1}l_{2}m_{2}}^{l_{3}m_{3}}$ 
is zero when $l_{1}+l_{2}+l_{3}=$even and 
is given as follows when $l_{1}+l_{2}+l_{3}=$odd: 

\begin{Eqnarray}
&&f_{l_{1}m_{1}l_{2}m_{2}}^{l_{3}m_{3}} \cr
&=&2\sqrt{(2l_{1}+1)(2l_{2}+1)(2l_{3}+1)}
\sqrt{N+1}(-1)^{N-1}
\left(\begin{array}{c c c}
  l_{1} & l_{2} & l_{3} \\ m_{1}  & m_{2} & m_{3} \\
 \end{array} \right)
\left\{ \begin{array}{c c c}
  l_{1} & l_{2} & l_{3} \\ \frac{N}{2}  & \frac{N}{2} & \frac{N}{2} \\
 \end{array} \right\}. 
\end{Eqnarray}

\noindent 
This quantity behaves as $N^{-1}$ when $N \rightarrow \infty$. 

%%%%%%%%%%%%%%%%%%%%%%%%%%%%%%%%%%%%%%%%%%%%%%%%%%%%%%%%
%%%%%%%%%%%%%%%%%%%%%%%%%%%%%%%%%%%%%%%%%%%%%%%%%%%%%%%%

\section{Wick type star product}
\setcounter{equation}{0}
\hspace{0.4cm}
Here we explain the star product on the fuzzy sphere which corresponds to 
a normal ordered operator product\cite{Bere, MMSZ, APS}
\footnote{
See \cite{SaWa, Ma} about Weyl ordered star
product on K\"{a}hler manifold.
} and 
derive mapping rules from matrix models to field theories with
the normal ordered products \`{a} la Berezin \cite{Bere}.

We first rescale $\hat{x}_i$ as
$\hat{y}_i=\hat{x}_i/\rho$ where $\hat{x}_i$'s are defined by
(\ref{clsol}). Then
\bango{
[\hat{y}_i,\hat{y}_j]=i\beta\epsilon_{ijk}\hat{y}_k,
}{a}
where $\beta=\da/\rho$ . Thus $\hat{y}_i^2=1$ and 
$\beta^2=4/N(N+2)$ . We define the stereographic projection
of $\hat{y}_i$ as
\bango{
\hat{z}=\frac{1}{\sqrt{2}}\hat{y}_-\hat{\chi}~,~\hat{z}^{\dagger}=
\frac{1}{\sqrt{2}}\hat{\chi}\hat{y}_+,
}{b}
where $\hat{y}_{\pm}=\hat{y}_1\pm i\hat{y}_{2}/\sqrt2$ ,
$\hat{\chi}=2(1-\hat{y}_3)^{-1}$ .
Note that
the complex coordinate projected from the sphere of radius $\rho$
is $\hat{w}=\rho\hat{z}$ . 
By the above definition, we obtain the commutation relation
\bango{
[\hat{z},\hat{z}^{\dagger}]=\beta\hat{\chi}\Big(1+|\hat{z}|^2
-\frac12\hat{\chi}(1+\frac{\beta}{2}|\hat{z}|^2)\Big),
}{c}
and by using $\hat{y}_i^2=1$ we can solve $\hat{\chi}$ 
in terms of $\hat{z}$ and $\hat{z}^{\dagger}$:
\bango{
\beta\hat{\chi}=2+\beta\hat{\xi}^{-1}-
\sqrt{4\hat{\xi}^{-1}+\beta^2\hat{\xi}^{-2}},
}{d}
where $|\hat{z}|^2=\hat{z}\hat{z}^{\dagger}$ and 
$\hat{\xi}=1+\beta|\hat{z}|^2$ .
For sufficiently large $N$, $\hat{\chi}\sim 1+|\hat{z}|^2$ and
the commutation relation is simplified as
\bango{
[\hat{z},\hat{z}^{\dagger}]\sim \frac1N(1+|\hat{z}|^2)^2.
}{d-1}

As discussed in \cite{MMSZ,APS,HP},
the operator $\hat{z}$ can be written in terms of 
an annihilation operator of a usual harmonic oscillator  as
\bango{
\hat{z}=f(\hat{n}+1)\hat{a},
}{d-2}
where
\bango{
[\hat{a},\hat{a}^{\dagger}]=1~,~\hat{n}=\hat{a}^{\dagger}\hat{a},
}{e}
and
\bango{
f(\hat{n})=\frac{\sqrt{N-\hat{n}+1}}{\sqrt{N/2(N/2+1)}+N/2-\hat{n}}
\sim \frac1{\sqrt{N+1-\hat{n}}}.
}{f}
Eq.(\ref{f}) is derived from a correspondence\cite{APS}
\bango{
\ket{N/2;m} \ \ \longleftrightarrow \ \ \ket{n}  ,
}{g-1}
where $\ket{N/2;m}$ and $\ket{n}$ are  eigenstates of $\hat{L}_3$ and $\hat{n}$
respectively. Thus the eigenvalue of $\hat{L}_3$ is given by
$m=n-N/2$ where  $0\le n\le N$ .
To construct a normal ordered star product, we define  the following
 coherent state $\ket{z}$. Requiring $\hat{z}\ket{z}=z\ket{z}$ , we obtain
\bango{
\ket{z}=M(|z|^2)^{-\frac12}\sum_{n=0}^{N}\frac{z^n}{[\gamma(n)]!}\ket{n},
}{g}
where $M(|z|^2)$ is a normalization factor, $\gamma(n)\equiv \sqrt{n}f(n)$ and
\[
[\gamma(n)]!=\ar{\{}{cl}{\gamma(n)\gamma(n-1)\cdots\gamma(1) & (n\ne 0)\\
1&(n=0)}{.}.
\]
Eq.(\ref{g}) satisfies the condition of the coherent state 
only for sufficiently large $N$ and when we can neglect the 
state of $| N + 1 \rangle $.
The normalization factor is determined from 
$\bk{z}{z}=1$ as
\bango{
M(|z|^2)=\sum_{n=0}^{N}\frac{|z|^{2n}}{[\gamma(n)^2]!}\sim
\sum_{n=0}^{N}[N+1-n]!\frac{|z|^{2n}}{n!}=(1+|z|^2)^{N}.
}{h}
Also the completeness condition
\bango{
1=\sum_{n}\ket{n}\bra{n}=(N+1)\int d\mu(z,\bar{z})\ket{z}\bra{z}
}{h-1}
determines the measure $d\mu$ as
\bango{
d\mu(z,\bar{z})\sim
\frac{idz\wedge d\bar{z}}{2\pi(1+|z|^2)^2}.
}{i}
(See \cite{APS} for details.)

Normal ordered operators can be expanded in terms of  $\ket{n}\bra{m}$ 
\bango{
\hat{Z}_{nm}=\ket{n}\bra{m}=\frac{(\hat{a}^{\dagger})^n}{\sqrt{n!}}\ket{0}
\bra{0}\frac{\hat{a}^m}{\sqrt{m!}}=\frac{(\hat{z}^{\dagger})^n}{[\gamma(n)]!}
\ket{0}\bra{0}\frac{\hat{z}^m}{[\gamma(m)]!},
}{j}
where $\ket{0}\bra{0}=\sum_kc_k(\hat{z}^{\dagger})^k\hat{z}^k$ and
$c_k$ is determined by $M(x)^{-1}=\sum_kc_kx^k$ .
The corresponding orthonormalized basis of functions are defined by
\bango{
Z_{nm}(z,\bar{z})=\bra{z}\hat{Z}_{nm}\ket{z}=
\frac{\bar{z}^nz^m}{[\gamma(n)]![\gamma(m)]!M(|z|^2)}.
}{o}
Thus any operators 
\bango{
\hat{a}=\sum_{nm}\tilde{a}_{nm}\hat{Z}_{nm}
}{o-1}
are mapped to corresponding functions as
\bango{
\hat{a}\to a(z,\bar{z})=\bra{z}\hat{a}\ket{z}
=\sum_{nm}\tilde{a}_{nm}Z_{nm}(z,\bar{z}).
}{k}
The trace of an operator is mapped to an integral over the projected plane
\bango{
\frac{1}{N+1}Tr\to\int d\mu(z,\bar{z}).
}{k-1}
The analytic continuation of this function is defined by
\bango{
a(z,\bar{\eta})=\frac{\bra{\eta}\hat{a}\ket{z}}{\bk{\eta}{z}},
}{l}
where $\bk{\eta}{z}=M(|\eta|^2)^{-\frac12}M(|z|^2)^{-\frac12}
M(\bar{\eta}z)$ .
Using this definition, 
a product of matrices is mapped to 
the star product as 
\eq{
a\star b(z,\bar{z}) &=& \bra{z}\hat{a}\hat{b}\ket{z}
=(N+1)\int d\mu(\eta,\bar{\eta})
\frac{\bra{z}\hat{a}\ket{\eta}}{\bk{z}{\eta}}|\bk{\eta}{z}|^2
\frac{\bra{\eta}\hat{b}\ket{z}}{\bk{\eta}{z}} \nn \\
&=& (N+1)\int d\mu(\eta,\bar{\eta})~a(\eta,\bar{z})
\frac{M(\bar{\eta}z)M(\bar{z}\eta)}{M(|\eta|^2)M(|z|^2)}b(z,\bar{\eta}) \nn \\
&\sim& (\frac1\lambda +1)
\int \frac{id\eta\wedge d\bar{\eta}}{2\pi(1+|\eta|^2)^2}
a(\eta,\bar{z})\Big[
\frac{(1+z\bar{\eta})(1+\eta\bar{z})}{(1+|\eta|^2)(1+|z|^2)}
\Big]^{\frac1\lambda}b(z,\bar{\eta}),
\label{m}}
where $\lambda=1/N$ .
This star product is nothing but the Berezin product on the sphere\cite{Bere}.
From the definition of $\hat{Z}_{nm}$, the corresponding function satisfies
the following simple relations
\bango{
Z_{nm}\star Z_{rs}=\delta_{mr}Z_{ns},
}{p}
and
\bango{
\int d\mu(z,\bar{z})\bar{Z}_{nm}\star Z_{n'm'}=\delta_{nn'}\delta_{mm'}.
}{r}

Next we write down functions corresponding to the operators $\hat{y}$ 's.
$\hat{y}_{\pm},\hat{y}_3$ are rewritten by $\hat{z},\hat{z}^{\dagger}$ and
$\hat{n}$ \footnote{$\hat{n}$ can be written in terms of 
$\hat{z},\hat{z}^{\dagger}$ . However it is not necessary here.}
as
\eq{
\hat{y}_3&=&\beta(\hat{n}-N/2), \nn \\
\sqrt2\hat{y}_+&=&(1-\hat{y}_3)\hat{z}^{\dagger}, \nn \\
\sqrt2\hat{y}_-&=&\hat{z}(1-\hat{y}_3).
\label{s}}
Thus, the corresponding functions can be expanded by $Z_{nm}$ as
\eq{
y_3&=&\beta\sum_{n=0}^{N}(n-N/2)Z_{nn}, \nn \\
y_+&=&(1-y_3)\star\bar{z}=\sum_{n=0}^{N-1}\tilde{y}_nZ_{n+1~n}, \nn \\
y_-&=&z\star(1-y_3)=\sum_{n=0}^{N-1}\tilde{y}_nZ_{n~n+1},
\label{t}}
where
\bango{
\tilde{y}_n=(1-\beta(n+1-N/2))\gamma(n+1)=\beta\sqrt{n+1}\sqrt{N-n}.
}{u}
Also the functions corresponding to the operators
$\hat{z},\hat{z}^{\dagger}$ can be expanded as
\eq{
z&=&\sum_{n=0}^{N-1}\gamma(n+1)Z_{n~n+1}, \nn \\
\bar{z}&=&\sum_{n=0}^{N-1}\gamma(n+1)Z_{n+1~n}.
\label{v}}

Now we will derive  differential operators corresponding to the
adjoint operators $Ad(\hat{y}_i)$ which are necessary to
rewrite the matrix model action in terms of field theory
on the projected plane.
% in order to rewrite the discuss  the conmmutative limit (\ref{comlim}). 
We first define differential operators
\eq{
k_z&=&\frac1NM^{-1}\rau_z M\sim \frac{\bar{z}}{1+|z|^2}+\frac1N\rau_z, \nn \\
k_{\bar{z}}&=&\frac1NM^{-1}\rau_{\bar{z}}M\sim\frac{z}{1+|z|^2}+
\frac1N\rau_{\bar{z}} .
\label{D}}
Note that the derivatives above act not only on $M$ but on functions
on the right of $M$. 
We can easily obtain the following relations
\eq{
zk_zZ_{nm}=m'Z_{nm}~&,&~\bar{z}k_{\bar{z}}Z_{nm}=n'Z_{nm}, \nn \\
k_{z}Z_{nm}=\frac{m'}{\gamma(m)}Z_{n~m-1}~&,&~
k_{\bar{z}}Z_{nm}=\frac{n'}{\gamma(n)}Z_{n-1~m}, \nn \\
zZ_{nm}=\gamma(m+1)Z_{n~m+1}~&,&~\bar{z}Z_{nm}=\gamma(n+1)Z_{n+1~m},
\nn}

\noindent 
where $n'=n/N~,~m'=m/N$ .
By using these relations, we have
\[
[\hat{y}_3,\hat{Z}_{nm}]\to\beta(n-m)Z_{nm}
=\beta N(\bar{z}k_{\bar{z}}-zk_z)Z_{nm}
\sim\beta(\bar{z}\rau_{\bar{z}}-z\rau_z)Z_{nm}
\]
and hence the adjoint operator in matrix models $Ad(\hat{y}_3)$
becomes the following differential operator after
mapping matrix models to field theories:
\bango{
Ad(\hat{y}_3)=[\hat{y}_3,\cdot]\to\beta(\bar{z}\rau_{\bar{z}}-z\rau_z) .
}{E}
We can also obtain similar rules for adjoint operators:
\eq{
Ad(\sqrt2\hat{y}_+)&\to& -\beta(\rau_z+\bar{z}^2\rau_{\bar{z}}),
\label{F} \\
Ad(\sqrt2\hat{y}_-)&\to&\beta(\rau_{\bar{z}}+z^2\rau_z),
\label{G} \\
Ad(\hat{y}_i)^2&\to& -\beta^2(1+|z|^2)^2\rau_z\rau_{\bar{z}}.
\label{H}}
Using mapping rules (\ref{k}),(\ref{k-1}),(\ref{m}) and
(\ref{F}) $\sim$ (\ref{H}), the matrix model actions can be 
written in terms of field theories on the projected plane.

In the commutative limit, the action is simplified
by writing the vector fields in terms of projected ones
as (\ref{killing}).
Killing vectors on the projected plane are given 
% we define local coordinates 
% In eq.(\ref{killing}), we reset $i=+,-,3$ and $a=z,\bar{z}$ . 
% Then we obtain
\eq{
K_+^z=-\frac{i}{\sqrt2}~&,
&~K_+^{\bar{z}}=-\frac{i}{\sqrt2}\bar{z}^2, \nn \\
K_-^z=\frac{i}{\sqrt2}z^2~&,
&~K_-^{\bar{z}}=\frac{i}{\sqrt2}, \nn \\
K_3^z=-iz~&,&~K_3^{\bar{z}}=i\bar{z}, \nn
}
which can be seen 
from (\ref{F})$\sim$(\ref{H}).
% where $\hat{L}_i=\hat{y}_i/\beta$ .
Thus the field strength of $U(1)$ gauge field in the commutative limit
can be written in terms of $z,\bar{z}$ as
\eq{
F_{+-}&=&-\frac{i}{\rho^2}K_+^aK_-^b(\rau_ab_b-\rau_bb_a)
=-\frac{i}{2\rho^2}(1-|z|^2)(1+|z|^2)F_{z\bar{z}}, \nn \\
F_{+3}&=&-\frac{i}{\rho^2}K_+^aK_3^b(\rau_ab_b-\rau_bb_a)
=-\frac{i}{\sqrt2\rho^2}\bar{z}(1+|z|^2)F_{z\bar{z}}, \nn \\
F_{-3}&=&-\frac{i}{\rho^2}K_-^aK_3^b(\rau_ab_b-\rau_bb_a)
=\frac{i}{\sqrt2\rho^2}z(1+|z|^2)F_{z\bar{z}},
\label{I}}
where 
\bango{
F_{z\bar{z}}=\rau_zb_{\bar{z}}-\rau_{\bar{z}}b_z.
}{J}
We have set the scalar field $\phi=0$ for simplicity. 
Also we get
\eq{
Tr \hat{F}^2&\to&2(N+1)\int d\mu(z,\bar{z})(F_{12}^2+F_{23}^2+F_{31}^2) \nn \\
&=& 2(N+1)\int d\mu(z,\bar{z})(-F_{+-}^2+2F_{+3}F_{-3}) \nn \\
&=& \frac{2(N+1)}{\rho^4}\int d\mu(z,\bar{z})(1+|z|^2)^4F_{z\bar{z}}^2 \nn \\
&=& \frac{16(N+1)}{\rho^4}\int d\mu(z,\bar{z})F_{ab}F^{ab},
\label{L}}
where
\bango{
g^{zz}=g^{\bar{z}\bar{z}}=0~,~g^{z\bar{z}}=g^{\bar{z}z}=\frac{(1+|z|^2)^2}{4}.
}{L-1}
This is the well-known result.

Finally we consider the flat limit (\ref{flat}). 
Here, it should not be discussed around the north pole but the south pole 
because of the definition (\ref{b}).
So $\hat{y}_3$ can be approximated as 
\bango{
\hat{y}_3\sim -\beta N/2\sim -1,
}{0}
and a rescaled coordinate corresponding to $x'$ is 
\bango{
w'=i\sqrt{\frac2N}w=i\sqrt{\frac2N}\rho z=i\rho'z.
}{1}
Thus we can rewrite the star product (\ref{m}) in the flat case:
\eq{
M(|z|^2)&\sim& (1+|z|^2)^{N} \nn \\
&=& (1-\frac{|w'|^2}{\rho'^2})^{N} \nn \\
&=& \Big((1-\frac{|w'|^2}{\rho'^2})^{-\frac{\rho'^2}{|w'|^2}}\Big)^{-\frac{N}{\rho'^2}|w'|^2} \nn \\
&\stackrel{\rho'\to\infty}{\longrightarrow}& e^{-\frac{N}{\rho'^2}|w'|^2},
\label{A}}
\eq{
d\mu(z,\bar{z})&\sim&
\frac{idz\wedge d\bar{z}}{2\pi(1+|z|^2)^2} \nn \\
&=&(1-\frac{|w'|^2}{\rho'^2})^{-2}\frac{-idw'
\wedge d\bar{w'}}{2\pi\rho'^2} \nn \\
&\stackrel{\rho'\to\infty}{\longrightarrow}& 
-\frac{idw'\wedge d\bar{w'}}{2\pi\rho'^2},
\label{B}}
\bango{
a\star b(w',\bar{w'})=\frac1{\nu}
\int \frac{id\eta^{\prime}\wedge d\bar{\eta^{\prime}}}
{2\pi}a(\eta^{\prime},\bar{w^{\prime}})
e^{\frac1{\nu}(w^{\prime}\bar{\eta^{\prime}}+
\eta^{\prime}\bar{w^{\prime}}-\eta^{\prime}\bar{\eta^{\prime}}
-w^{\prime}\bar{w^{\prime}})}b(w^{\prime},\bar{\eta^{\prime}}),
}{C}
where $\nu=-\rho'^2/(N+1)\sim -\da^2/2$ .
Thus the star product becomes the Berezin product on the flat plane.

%%%%%%%%%%%%%%%%%%%%%%%%%%%%%%%%%%%%%%%%%%%%%%%%%%%%%%%%%%%%%%%%%%%%%%
%%%%%%%%%%%%%%%%%%%%%%%%%%%%%%%%%%%%%%%%%%%%%%%%%%%%%%%%%%%%%%%%%%%%%%

\end{document}